\documentclass[reqno]{amsart}

\usepackage{xcolor}
\usepackage{mathtools}
\usepackage{amsthm}
\usepackage{graphicx}
\usepackage{enumerate}
\usepackage{mathbbol}
\usepackage{bbm}
\usepackage{dashrule}
\usepackage{xfrac}
\usepackage{cases}
\usepackage{makecell, booktabs, longtable}
\usepackage[colorlinks=true, allcolors=blue]{hyperref}

\DeclarePairedDelimiter{\diagfences}{(}{)}
\newcommand{\diag}{\operatorname{diag}\diagfences}

\newcommand{\bq}{\bar{q}}

\newcommand{\Eof}[1]{\displaystyle\mathbb{E}\left[ #1 \right]}
\newcommand{\QV}[1]{\displaystyle\mathbb{QV}\left[ #1 \right]}
\newcommand{\Vad}{\mathcal{V}_{\text{ad}}}

\newcommand{\bs}{\boldsymbol}
\newcommand{\tr}{\mathrm{tr}}

\newcommand{\R}{\mathbb{R}}
\newcommand{\bA}{\boldsymbol{A}}
\newcommand{\bG}{\boldsymbol{G}}
\newcommand{\bR}{\boldsymbol{R}}
\newcommand{\bX}{\boldsymbol{X}}
\newcommand{\ba}{\boldsymbol{a}}
\newcommand{\bb}{\boldsymbol{b}}
\newcommand{\be}{\boldsymbol{e}}
\newcommand{\bk}{\boldsymbol{k}}
\newcommand{\br}{\boldsymbol{r}}
\newcommand{\bSigma}{\boldsymbol{\Sigma}}
\newcommand{\tM}{\widetilde M}
\newcommand{\talpha}{\widetilde\alpha}
\newcommand{\teta}{\widetilde\eta}

\newcommand{\kws}{\bigskip\par\addvspace\medskipamount{\rightskip=0pt plus1cm}{\noindent \bfseries Keywords: \enspace}}

\newcommand{\JEL}{\bigskip\par\addvspace\medskipamount{\rightskip=0pt plus1cm}{\noindent \bfseries JEL Classification: \enspace}}

\definecolor{capri}{rgb}{0.0, 0.75, 1.0}

\theoremstyle{plain}
\newtheorem{theorem}{Theorem}[section]
\newtheorem{definition}[theorem]{Definition}
\newtheorem{lemma}[theorem]{Lemma}
\newtheorem{proposition}[theorem]{Proposition}
\newtheorem{remark}[theorem]{Remark}
\newtheorem{corollary}[theorem]{Corollary}

\newtheorem{problem}[theorem]{Problem}

\usepackage[natbib,hyperref,useprefix,bibstyle=authoryear,citestyle=authoryear-comp,maxbibnames=99,backend=bibtex,dashed=false]{biblatex} \usepackage{xpatch}
\xpatchbibmacro{volume+number+eid}{\setunit*{\adddot}}{}{}{} \DeclareFieldFormat[article]{number}{\mkbibparens{#1}}
\renewbibmacro{in:}{}

\bibliography{biblio-ANOR}

\usepackage{hyperref}

\begin{document}

\allowdisplaybreaks

\title[Optimal order execution under price impact: A hybrid model]
{Optimal order execution under price impact: \\ A hybrid model}

\author[M. Di Giacinto, C. Tebaldi, and T.-H. Wang]
{Marina Di Giacinto, Claudio Tebaldi, and Tai-Ho Wang}

\address{Marina Di Giacinto \newline
Dipartimento di Economia e Giurisprudenza \newline
Universit{\`a} degli studi di Cassino e del Lazio Meridionale, Cassino (Fr), Italy}
\email{digiacinto@unicas.it}

\address{Claudio Tebaldi \newline
Università Commerciale Luigi Bocconi {\rm and} IGIER {\rm and} \text{Baffi-Carefin}\newline
Milano, Italy}
\email{claudio.tebaldi@unibocconi.it}

\address{Tai-Ho Wang \newline
Department of Mathematics \newline
Baruch College, The City University of New York \newline
1 Bernard Baruch Way, New York, NY10010 \newline
{\rm and} \newline
China Center for Economics Research (CCER), Peking University \newline
Beijing, China
}
\email{tai-ho.wang@baruch.cuny.edu}

\begin{abstract}
In this paper we explore optimal liquidation in a market populated by a number of heterogeneous market makers that have limited inventory-carrying and risk-bearing capacity. We derive a reduced form model for the dynamic of their aggregated inventory considering a proper scaling limit.  The resulting price impact profile is shown to depend on the characteristics and relative importance of their inventories. The model is flexible enough to reproduce the empirically documented power law behavior of the price impact function. For any choice of the market makers characteristics, optimal execution within this modeling approach can be recast as a linear-quadratic stochastic control problem in which the value function and the associated optimal trading rate can be obtained semi-explicitly subject to solving a differential matrix Riccati equation. Numerical simulations are conducted to illustrate the performance of the resulting optimal liquidation strategy in relation to standard benchmarks. Remarkably, they show that the increase in performance is determined by a substantial reduction of higher order moment risk.
\end{abstract}

\maketitle

\kws Optimal execution, Price impact, Inventory cost, Order fill uncertainty, Stochastic optimal control.

\JEL C61, G10, G11.


\tableofcontents


\section{Introduction}

Price formation in real markets is the outcome from a complex interaction between traders with heterogenous trading technologies and specialized intermediaries that promote exchange by providing liquidity at a cost. In the traditional no arbitrage pricing framework, prices are completely inelastic to demand shocks so that microstructure of price formation is irrelevant. This modeling approach has a number of important drawbacks. First, it is unable to discriminate among order execution algorithms. Second, as an implication, any information on the relationship between transacted quantities and price variations is completely missing.

A key role in shaping the (possibly optimal) mechanics of order execution is played by the so called price impact function. It describes how the execution of an order impacts the transacted prices during its course  of execution. There's a vast literature analyzing its empirical shape, see for instance \textcite{Bac-at-al} and the references therein, proving that it increases as a power law during execution and, when the execution is complete, it reverts back to certain level, usually differing from the price
at the inception.

This empirical fact is usually not encoded within the conventional modeling approaches that are in use to assess the best execution strategies. For example, in the Almgren-Chriss model the impact profile is a time-weighted-average-price (TWAP) execution which is inconsistent with empirical observations in that the profile is linear and it shows no price reversion after completion. Vice-versa, traditional models known under the name of transient impact models, produce an impact curve that misses to reproduce the permanent impact effect.

Empirical analysis has also highlighted that market makers post two-way prices to buy or sell inventories to maximize their expected utility of wealth. Liquidity suppliers and market markers profit from providing immediacy to less patient investors, but have limited inventory-carrying and risk-bearing capacity. Their main risk exposure consists of imbalances in proceeds and inventories accumulated through transactions that might eventually be realized in presence of adverse price movements.

This implies that it is possible to put the price gap between mid-prices and transacted prices in direct connection with the inventory costs of a spectrum of market makers. This price gap arguably is determined by following major components: temporary, permanent and transitory price impact, and processing costs. In this paper, we jointly model temporary, permanent and transitory price impact cost components. Processing costs are neglected since it mostly consists of fee or rebate and has almost no
price impact.

We consider a trader whose goal is to design an order execution schedule to optimally liquidate a position under the price impact model. We assume that the trader considers as performance criterion the risk-adjusted profit-and-loss (P\&L) plus a\footnote{
Negation of the transaction P\&L is equivalent to what is usually referred to as the implementation shortfall in the trading and order execution industry/community, see Section \ref{sec:pnl}.} risk component that is given by the quadratic variation of the P\&L process during execution. Finally, the risk-adjusted P\&L is further penalized by a term incurred from a final block trade at the acquisition/liquidation horizon, should there be a block trade required to fully close the position.

More recently, \textcite{GraHor} analyze optimal liquidation inmarkets where both instantaneous and transitory price impact when only absolutely continuous trading strategies are admissibl while \textcite{NeuVos} analyze the liquidation problem in  the presence of linear temporary and transient price impact and price predictability

The novel aspect of this paper relies on the fact that we explicitly take into account the transitory impact on prices of a multiplicity of market makers' inventories. In particular, we apply a proper scaling procedure to the optimal inventory dynamics found by \textcite{GueLehFer} and derive a reduced form, in the so-called hydrodynamic limit, for the market maker's optimal inventory dynamics. The reduced form model reproduces the large scale behavior of the original model and consists of a simple mean reverting process that is therefore characterized uniquely by three parameters, mean reversion, long term rate and volatility. In practice, the transitory component of the price process is further split into the sum of contributions from netted inventories of individual market makers.

Then we adopt this reduced form description and compute the shape of the price impact function in terms of the aggregated inventory of a crowd of market makers with heterogeneous degree of mean reversion under the assumption that there's a term directly related to the global volume imbalance that affects directly the price dynamics.

The resulting price impact profile is shaped by the distribution of market maker's mean reversions. The resulting model has sufficient flexibility to reproduce the empirical shape of the observed price-impact functions while preserving analytical tractability of the resulting optimal liquidation problem.\footnote{
Whether this representation result is a mathematical artifact or empirically relevant is a question that can be answered only on a purely empirical basis and is left for future research.}

For any market configuration, the computation of the optimal execution policy is reduced to a (non-standard) linear-quadratic problem and thus is analytically solvable.

In particular, following a strategy similar to the one put forward in \textcite{bouchaud2003fluctuations} and considering a model with a proper distribution of market makers, the proposed modelling approach can accomodate the power-law shape of the impact function that has been empirically detected and generates a closed form solution for the corresponding trader's optimal liquidation policy.

In the numerical section of the paper we run a comparative analysis among liquidation policies. We show that the optimal strategy improves the performance over more traditional ones since it can take properly into account also higher order moment risk. Simulations show that a large reduction in costs arising from variance, skewness and higher order moment risk components can be achieved accepting only a slight increase in the mean
expected cost. The most efficient tradeoff is captured by the model thanks to the presence of two key modeling features. First, the investor's position dynamics includes a small dynamic uncertainty component. Second, the performance indicator includes quadratic variation component as risk. Then the numerical simulation illustrates how the resulting optimal liquidation strategy takes properly into account the uncertain nature of the trading outcome and optimizes trader's action accordingly. The rest of the paper is organized as follows. We frame the model underlying the optimal execution problem under price impact in Section \ref{sec:model-setup}. Section \ref{sec:control-problem} recasts the optimal order execution problem as an LQ stochastic control problem and summarizes its solution and the associated optimal trading strategies. Technical proofs and verification theorems are collected in Section \ref{sec:verification}. Numerical illustrations and discussion of the results are reported in Section \ref{sec:numerical}. Section \ref{sec:conclusion} concludes the paper.


\section{Model setup} \label{sec:model-setup}

Throughout the paper, $\left(\Omega, \mathcal{F}, \mathbb{P}\right)$ denotes
a complete probability space equipped with a filtration describing the
information structure $\mathbb{F} := \left \{ \mathcal{F}_{t}\right \}_{t
\in [0,T]}$ -- where $t$ is the time variable and $T>0$ the fixed finite
liquidation horizon, $\left \{B_S(t), B_Q(t), B_X(t)\right \}_{t \in [0,T]}$ is a three-dimensional uncorrelated Brownian motion defined on $\left(\Omega, \mathcal{F}, \mathbb{P}\right)$, and $\mathbb{F}$ is the filtration generated by the trajectories of the above Brownian motion, completed with
all $\mathbb{P}$-null measure sets of $\mathcal{F}$.

In a continuous time setting, we outline the model and describe the problem
faced by an investor when liquidating a given amount of shares of certain
stock within the time interval $[0,T]$, in a market where liquidity provision is operated by market makers facing
inventory risk, i.e., the risk related to the signed quantity of shares they
hold.


\subsection{Reduced form inventory dynamics} \label{sec:inventory}

In this subsection we briefly introduce the market
maker's optimal management inventory problem as derived in \textcite{GueLehFer}. Then we introduce the reduced form of market makers'
inventory dynamics that will be considered in this paper.

A market maker provides liquidity by posting two-way prices at which she is
willing to buy or sell, i.e., the bid and ask prices respectively.  A market maker faces
the risk stemming from the uncertainty in the value of their holdings in the
asset, this is the so called inventory risk. It was first examined theoretically in \textcite{Gar}, \textcite{Sto}, \textcite{AmiMen}, \textcite{HoSto}, and more recently in \textcite{AveSto}
as well as \textcite{GueLehFer}. We briefly introduce the framework
set up by \textcite{AveSto} and in \textcite{GueLehFer}, and state the
elegant result obtained in \textcite[p.~487]{GueLehFer}.

Denote by $S^{b}$ and $S^{a}$ the bid and ask prices, respectively, and by $\delta ^{b}$ and $\delta ^{a}$ the difference between the quotes and the
reference price $S$, i.e., $\delta ^{b}=:S-S^{b}$ and $\delta ^{a}=:S^{a}-S$. The market maker's inventory problem is to continuously quote the two
prices $S^{a}$ and $S^{b}$, or equivalently to determine the spreads $\delta^{a}$ and $\delta ^{b}$ respectively, in order to maximize their expected
utility of wealth which consists of proceeds and inventories accumulated
through transactions with other investors in the market within either a
finite or infinite horizon. In the works of \textcite{AveSto} and \textcite{GueLehFer}, the reference price $S$ is assumed following an
arithmetic Brownian motion, driftless or constant drift. The arrival of
market buy and market sell orders are assumed independent Poisson processes.
A market maker's objective is to maximize their expected utility of terminal
wealth consisting of proceeds and inventory throughout a market making
horizon. By further assuming exponential utility $U$ given by
\begin{equation*}
U(w)=\frac{1}{\nu }\left( 1-e^{-\nu w}\right)
\end{equation*}
and exponential arrival rates $\theta _{a}$ and $\theta _{b}$ for market buy
and sell orders as functions of $\delta ^{a}$ and $\delta ^{b}$,
respectively, given by
\begin{equation}
\theta _{a}(\delta ^{a})=Ae^{-\kappa \delta ^{a}},\qquad \theta _{b}(\delta
^{b})=Ae^{-\kappa \delta ^{b}},  \label{eq:arrival-rates}
\end{equation}
for some constants $A$ and $\kappa $, \textcite{GueLehFer} obtained the
following elegant approximate optimal spreads $\delta ^{b}$ and $\delta ^{a}$
\begin{subequations}
\label{eq:delta}
\begin{align}
\delta ^{b}& \approx \frac{1}{\nu }\ln \left( 1+\frac{\nu }{\kappa }\right)
+\left( q+\frac{1}{2}-\frac{\mu }{\nu \sigma ^{2}}\right) \sqrt{\frac{\sigma
^{2}\nu }{2\kappa A}\left( 1+\frac{\nu }{\kappa }\right) ^{1+\frac{\kappa }{\nu}}},  \label{eq:delta-b} \\
\delta ^{a}& \approx \frac{1}{\nu }\ln \left( 1+\frac{\nu }{\kappa }\right)
+\left( -q+\frac{1}{2}+\frac{\mu }{\nu \sigma ^{2}}\right) \sqrt{\frac{\sigma ^{2}\nu }{2\kappa A}\left( 1+\frac{\nu }{\kappa }\right) ^{1+\frac{\kappa }{\nu }}},
\end{align}
\end{subequations}
where $\mu $ and $\sigma $ denote respectively the drift and volatility of $S$, $q$ the market maker's current inventory.

We are now ready to state our first result:

\begin{proposition}
Consider the quoting rules given in \eqref{eq:delta}. Then scaling the
parameter $A$ by $\frac{A}{h^{2}}$ in the arrival rates, as well as scaling
down the Poisson processes $N^{b}$ and $N^{a}$ to $hN^{b}$ and $hN^{a}$
respectively, then in the limit as $h\rightarrow 0$, the evolution of
market maker's inventory converges to an Ornstein-Uhlenbeck process:

\begin{equation*}
\begin{cases}
dQ(t)=\theta (\bq^{0}-Q_{i}(t))dt+\sigma _{Q}dB_{Q}(t),\quad t\in \lbrack
0,T] \\
Q(0)=0
\end{cases}
\end{equation*}
{\normalsize with mean reversion $\theta =2c_{1}c_{2}\kappa $, long term
mean }$\bq^{0}=${\normalsize $\frac{\mu }{\nu \sigma _{S}^{2}}$, and
volatility }$\sigma _{Q}=${\normalsize $2\sqrt{c_{1}}$.}
\end{proposition}

\begin{proof}
See Appendix \ref{app:conv-ou}.
\end{proof}

This rescaling limit, the so-called
hydrodynamic limiting procedure, considers the
optimal inventory dynamics in the limit $h\rightarrow 0$ assuming that in this limit the the arrival rate
diverges but the marginal impact of each transaction becomes negligible.

The resulting reduced form dynamics is the one relevant for the trader that is missioned to liquidate a large
position. The scaling procedure removes any reference to the micro-structure mechanics of the order execution. For example, this is the situation faced by a trader that delegates the optimization of the order selection to a high-frequency specialized intermediary or to an algorithmic procedure.

The second important assumption we introduce is that the market is populated by heterogeneous market makers, each one
characterized by the same reduced form dynamics but different characteristics. The dynamics of the $i-$th market maker's inventory $Q_{i}$ will follow an Ornstein-Uhlenbeck process:
\begin{equation}
\begin{cases}
dQ_{i}(t)=\theta _{i}(\bq_{i}(t)-Q_{i}(t))dt+\sigma _{Q_{i}}dB_{Q}(t),\quad
t\in \lbrack 0,T] \\
Q_{i}(0)=0
\end{cases}
\qquad \mbox{ for }i=1,2,\dots ,n,  \label{eq:inventory}
\end{equation}
where

\begin{itemize}
\item The initial inventory $Q_i(0)$, $i= 1, \dots, n$, is assumed 0 for simplicity;

\item All the $Q_{i}$'s are driven by the same Brownian motion $B_{Q}$.

\item The capacity of the $i^{\text{th}}$ market maker is proxied by $\bar{q}_{i}(t)=\bq_{i}^{1}v(t)+\bq_{i}^{0}$, assuming $\bq_{i}^{1}$ and $\bq_{i}^{0}
$ constants.
\end{itemize}

While the reduced form derived in the previsous Proposition corresponds
to $\bq_{i}^{1}=0$, in the model formulation we consider also the possibility of a
feedback, i.e. that the trading rate of market maker $i$ is selected based on
the agent's trading rate $v\colon \Omega \times \lbrack 0,T]\rightarrow \mathbb{R}$ which will be regarded as the main control variable in the formulation of the agent's liquidation problem.

The linear coefficient $\bq_{i}^{1}$ reflects the $i^{\text{th}}$ market
maker's direct reaction to the investor's liquidation rate, whereas the
constant term $\bq_{i}^{0}$ can be interpreted as an upfront capacity set up
by the $i^{\text{th}}$ market maker as a passive attempt to maintain their
inventory close to a fixed capacity level during execution. Upon completion
of order execution, they all revert their inventory back to zero.

The relative importance of each market makes in the market is set by a
weight $\nu _{i}>0$, $i=1,\dots ,n$, such that $\sum_{i=1}^{n}\nu _{i}=1$.
Then, we can define a netted aggregate market volume of orders:
\begin{equation*}
Q^{M}(\cdot ):=\sum_{i=1}^{n}\nu _{i}Q_{i}(\cdot ).
\end{equation*}
Every transaction (buy or sell) is assumed traded with any of the market
makers. Then $Q^{M}$ will have a dynamics given by the following equation
\begin{equation*}
\begin{cases}
dQ^{M}(t)=\sum_{i=1}^{n}\nu _{i}\theta _{i}(\bq_{i}(t)-Q_{i}(t))dt+\sigma
_{Q}^{M}dB_{Q}(t),\quad t\in \lbrack 0,T], \\
Q^{M}(0)=0,
\end{cases}
\end{equation*}
with $\sigma _{Q}^{M}:=\sum_{i=1}^{n}\nu _{i}\sigma _{Q_{i}}$.


\subsection{Investor's trading strategy}

As in \textcite{QF,JORS}, the evolution of the investor's position $X$ is assumed to
satisfy the following stochastic differential equation
\begin{equation}
\begin{cases}
dX(t)=-v(t)dt+mdB_{X}(t),\quad t\in \lbrack 0,T], \\
X(0)=x_{0}>0,
\end{cases}
\label{eq:position}
\end{equation}
where $m\geq 0$ measures the magnitude of the uncertainty, while the quantity $x_{0}$ represents the initial position to be liquidated by $T$. The non-controlled diffusive component in the dynamics of the investor's position $X$ in \eqref{eq:position}, takes into account that in real situations the agent's wealth process $X$ includes a small uncertainty component. The recent paper \textcite{CarLea}, see also \textcite{CarWeb}, produces compelling econometric evidence about the existence of a non-zero quadratic variation component in the time series of
various institutional investors' positions of Toronto Stock-Exchange that are publicly available. From a modeling point of view such term is well characterized: it represents the uncertainty that affects also quantities, i.e. the inventory of the trader, beyond prices. Note that the trader can only control the rate of trading, i.e., the drift of the process. Hence introducing this term, the optimal liquidation strategy will also take into account the risk arising from the quantity uncertainty. Traditional model dynamics are recovered in the limit of $m\rightarrow 0$.

While it would be analytically feasible to consider a non-zero level of correlation
between price and quantities, we follow \textcite{CarLea} (see Remark 1 of
that paper) and solve the model for the zero correlation case our modeling
approach does not explicitly model the price formation mechanics. In fact,
as discussed in \textcite{CarWeb} a limit order book execution would imply a
positive correlation while a market order would imply a negative
correlation. As previosuly stated, we assume a reduced form expression of
inventory dynamics that abstracts from the effective market microstructure
of the order execution.


\subsection{Traded price dynamics} \label{sec:price-evolution}

The dynamic evolution of the fair price $S$ of the stock is assumed to be
\begin{equation*}
\begin{cases}
dS(t)=\gamma dX(t)-\phi dQ^{M}(t)+\mu dt+\sigma _{S}dB_{S}(t),\quad t\in
\lbrack 0,T], \\
S(0)=s_{0}>0,
\end{cases}
\end{equation*}
namely, in integral form,
\begin{equation}
S(t)=s_{0}+\mu t+\sigma _{S}B_{S}(t)+\gamma \left( X(t)-x_{0}\right) -\phi
Q^{M}(t),\quad t\in \lbrack 0,T].  \label{eq:S-integral-form}
\end{equation}

In other words, the fair price $S$ is driven by an Arithmetic Brownian
motion with drift equal to $\mu \geq 0$ and volatility $\sigma _{S}>0$,
along with an inventory cost equal to $\phi \geq 0$ and a linear permanent
impact with parameter $\gamma \geq 0$. More specifically, the permanent
impact is modeled taking into account the continuous version of \textcite{AlmChr} model, whereas the inventory cost is framed following \textcite[Subsection~5.2,p.~487]{GueLehFer}: the term $\phi Q^{M}(t)$ quantifies the price pressure determined by the total
inventories carried by the market makers at time $t$.

Taking into account the dynamics of the position $X$ and the inventory $Q^{M}
$ as given by \eqref{eq:inventory} and \eqref{eq:position}, respectively,
the dynamics of the fair value price $S$ can be written as
\begin{equation*}
\begin{cases}
\begin{aligned} dS(t)=\left[\mu -\phi{\sum_{i=1}^{n}} \nu_i
\theta_{i}(\bq_{i}(t) - Q_{i}(t)) -\gamma v(t)\right]dt -\phi
\sigma^{M}_{Q}dB_{Q}(t) +\gamma m dB_{X}(t) +\sigma_{S} dB_{S}(t),\\ t\in
[0,T], \end{aligned} \\
S(0)=s_{0}>0,
\end{cases}
\end{equation*}
or in integral form as
\begin{equation*}
S(t)=s_{0}+\mu t-\int_{0}^{t}\bigg[\gamma v(u)+\phi \sum_{i=1}^{n}\nu
_{i}\theta _{i}\left( \bq_{i}(t)-Q_{i}(u)\right) \bigg]du-\phi \sigma
_{Q}^{M}B_{Q}(t)+\gamma mB_{X}(t)+\sigma _{S}B_{S}(t).
\end{equation*}

Following \textcite{AlmChr}, the transacted price $\widetilde{S}$ consists
of the fair price and a slippage referred as temporary impact
\begin{equation}
\widetilde{S}(t)=S(t)-\eta v(t),\quad t\in \lbrack 0,T],  \label{eq:S-tilde}
\end{equation}
that is, the transacted price reflects a temporary impact given by a linear
function of the current trading rate of market order $v$ with size $\eta >0$
. Note that $\eta v(t)$ is regarded as the informational cost, at time $t$
such as in the \textcite{Kyl} model. From the above relations, we can derive
the following transacted price expression
\begin{multline*}
\widetilde{S}(t)=s_{0}+\mu t+\sigma _{S}B_{S}(t)-\eta v(t)-\gamma
\int_{0}^{t}v(u)du+\gamma mB_{X}(t) \\
-\phi \int_{0}^{t}\sum_{i=1}^{n}\nu _{i}\theta _{i}\left( \bq_{i}-Q_{i}(u)\right) du-\phi \sigma _{Q}^{M}B_{Q}(t),\quad t\in \lbrack 0,T]
\end{multline*}

The transient
contribution determined as a function of the aggregate volume as determined by the population of market makers changes the shape of the price-impact
function.  In the case of a single market maker, it is easy to verify that the resulting
price-impact curve has an exponential shape. The composition of multiple modulated exponential terms offers a
natural approximation tool that may be used to reproduce a more realistic, power law shape of the price impact function. This 'heterogenous agents' approach that produces a power law response is well-know by now. It has been discussed in the context of volatility models by \textcite{andersen1997heterogeneous} and in the context of a propagator model in \textcite{bouchaud2003fluctuations}.  See also \textcite{OTTB} and the references inside for a review of a similar decomposition which has been introduced and is used in appplied econometric analysis of discrete time series.


\subsection{Profit and loss} \label{sec:pnl}

Following \textcite[Section 2.4, p.~10]{AlmChr}, the profit and loss (P\&L) $\Pi^{0}(t)$ of a trading strategy earned over the time interval $[0,t]$, $t\leq T$, is defined as
\begin{equation}  \label{eq:P&L}
\Pi^{0}(t) := X(t)\left( S(t)-S(0)\right) +\int_{0}^{t} (S(0)-\widetilde{S}
(u)) dX(u).
\end{equation}

The P\&L defined above can be decomposed in a self-financing strategy
contribution and a slippage component. They generalize the usual
self-financing relationships of frictionless markets to make it compatible
with markets with frictions, including the presence of the uncertainty in
the agent inventory as defined in our model. More details about the
decomposition of the P\&L formula can be found in
\textcite[Appendix
A.1~pp.~1673--1674]{JORS}.

Taking into account \eqref{eq:inventory}, \eqref{eq:position} and \eqref{eq:S-tilde}, the P\&L over the time interval $[0,t]$ can be rewritten
as
\begin{equation*}
\Pi^{0}(t)=X(t)S(t)-x_{0}s_{0}-\int_{0}^{t}S(u)dX(u)+\int_{0}^{t}\eta
v(u)dX(u)
\end{equation*}
Furthermore, for any given trading strategy $v(\cdot )$, the P\&L can be
calculated explicitly as
\begin{multline*}
\Pi^{0}(t)=\frac{\gamma }{2}\left( X^{2}(t)-x_{0}^{2}\right) +\frac{\gamma
}{2}m^{2}t-\phi X(t)Q^{M}(t) \\
+\int_{0}^{t}\left( -\eta v^{2}(u)-\phi Q^{M}(u)v(u)+\mu X(u)\right) du \\
+m\int_{0}^{t}\left( \eta v(u)+\phi Q^{M}(u)\right) dB_{X}(u)+\sigma
_{S}\int_{0}^{t}X(u)dB_{S}(u),
\end{multline*}
or, equivalently, by applying the integration by parts formula
\begin{multline}
\Pi^{0}(t)=\gamma m^{2}t+\int_{0}^{t}\bigg \{-\eta v^{2}(u)-\gamma v(u)X(u) +\bigg[\mu -\phi \sum_{i=1}^{n}\theta _{i}\left( \bar{q}_{i}-Q_{i}(u)\right) \bigg]X(u)\bigg \}du  \label{eq:P&L-to-be-max-m0-M} \\
-\int_{0}^{t}\phi \sigma _{Q}^{M}X(u)dB_{Q}(u)+m\int_{0}^{T}\left( \eta
v(u)+\gamma X(u)\right) dB_{X}(u)+\sigma _{S}\int_{0}^{t}X(u)dB_{S}(u).
\end{multline}


\subsubsection{Penalty of final block trade}

Following \textcite{QF, JORS}, we take into consideration as a penalty for a
final block trade of size $x \in \mathbb{R}$ a quadratic function
\begin{equation}  \label{eq:penalty-function}
f(x) := \beta x^{2}, \qquad \beta > 0.
\end{equation}
Thus, the P\&L posterior to the final block trade is given by
\begin{equation*}
\Pi^{0}(T) - \beta X^{2}(T),
\end{equation*}
should there be $X(T)$ shares of the stock remaining to be executed at the
horizon $T$.


\subsubsection{Quadratic variation}

We also choose to penalize the expected P\&L by its quadratic variation as
in \textcite{For-et-al}. From \eqref{eq:P&L-to-be-max-m0-M} the
instantaneous quadratic variation $\QV{\Pi^{0}(t)}$ of P\&L $\Pi^{0}(t)$ at
time $t \in [0,T]$, is given by
\begin{equation}  \label{eq:quadratic-variation}
\begin{aligned} d\QV{\Pi^{0}(t)} &= \phi^{2} (\sigma^{M}_{Q})^{2} X^{2}(t)
dt + m^{2} \left(\eta v(t) +\gamma X(t) \right)^{2} dt + \sigma_{S}^{2}
X^{2}(t) dt \\ &= \left[m^{2} \eta^{2} v^{2}(t) + 2 m^{2} \eta \gamma X(t)
v(t) + \left(\phi^{2} (\sigma^{M}_{Q})^{2} + m^{2} \gamma^{2} +
\sigma^{2}_{S} \right) X^{2}(t) \right] dt. \end{aligned}
\end{equation}

Therefore, the P\&L posterior to the final block trade and penalized by its
quadratic variation for some risk aversion parameter $\lambda \geq 0$ reads
\begin{equation}  \label{eq:P&L-QV-1}
\begin{aligned} &\Pi^{0}(T) -\beta X^{2}(T) -\lambda \QV{\Pi^{0}(T)} \\ &=
-\beta X^{2}(T) + \frac{\gamma}{2}\left(X^{2}(T) -x_{0}^{2}\right) +
\frac{\gamma}{2}m^{2} T - \phi X(T)Q^{M}(T) \\ &\quad +\int_{0}^T
\left(-\eta v^{2}(u) - \phi Q^{M}(u)v(u) +\mu X(u) \right) du \\ &\quad +m
\int_{0}^T \left(\eta v(u) + \phi Q^{M}(u)\right) dB_{X}(u)
+\sigma_{S}\int_{0}^{T} X(u) dB_{S}(u) \\ &\quad -\lambda \int_{0}^{T}
\left[m^{2} \eta^{2} v^{2}(t) +2 m^{2} \eta \gamma X(t) v(t) +\left(\phi^{2}
(\sigma^{M}_{Q})^{2} + m^{2} \gamma^{2} + \sigma^{2}_{S} \right)
X^{2}(t)\right]dt \\ &= \gamma m^{2}T - \frac \gamma2 x_0^2 + \left(\frac
\gamma2 - \beta \right) X^{2}(T) - \phi X(T)Q^{M}(T) \\ &\quad +\int_{0}^T
\left \{ -\widetilde \eta v^2_t - \phi Q^{M}_t v_t - \widetilde \xi X_t v_t
- \psi X_t^2 + \mu X_t \right \} dt \\ &\quad +m \int_{0}^T \left(\eta v(u)
+ \phi Q^{M}(u)\right) dB_{X}(u) + \sigma_{S} \int_{0}^{T} X(u) dB_{S}(u)
\end{aligned}
\end{equation}
where
\begin{subequations}
\label{eq:parameter-P&L-QV}
\begin{align}  \label{eq:psi}
\widetilde{\eta} &:= \eta(1 + \lambda m^{2} \eta), \\
\widetilde{\xi} & := 2\gamma \lambda m^{2}\eta, \\
\psi &:= \lambda \left( \phi^{2}(\sigma_Q^{M})^{2} + m^{2} \gamma^{2} +
\sigma^{2}_{S} \right).
\end{align}
\end{subequations}


\section{The optimal control problem} \label{sec:control-problem}

The order execution problem can be naturally recast as a linear quadratic (LQ) stochastic control problem. To this end, denote by $\bs A = \diag{-\theta_{1}, \cdots, -\theta_{n}, 0}$ the $(n+1)\times (n+1)$  diagonal matrix whose diagonal entries starting in the upper left corner being $-\theta_{1}, \dots, -\theta_{n}, 0$, and

\small
\begin{equation*}
\bX := \left[\begin{array}{c}
Q_1 \\
\vdots \\
Q_n \\
X
\end{array}\right], \,
\ba := \left[\begin{array}{c}
\theta_1 \bar q_1^1 \\
\vdots \\
\theta_n \bar q_n^1 \\
-1
\end{array}\right], \,
\bb := \left[\begin{array}{c}
\theta_1 \bar q_1^0 \\
\vdots \\
\theta_n \bar q_n^0 \\
0
\end{array}\right], \,
\bSigma := \left[\begin{array}{cc}
\sigma_{Q_{1}} & 0  \\
\sigma_{Q_{2}} & 0 \\
\vdots & \vdots \\
\sigma_{Q_{n}} & 0 \\
0 & m
\end{array}\right],\\
\bs{B} := \left[\begin{array}{c}
B_{Q} \\
B_{X}
\end{array}\right], \,
\bs{x} := \left[\begin{array}{c}
0  \\
0 \\
\vdots \\
x
\end{array}\right].
\end{equation*}
\normalsize
The state equation is governed by the following non-homogenous linear stochastic differential equation
\begin{equation} \label{eq:SDE-m0}
\begin{cases}
d\bs{X}(u) = \left(\bs{A}\bs{X}(u) + \bs{a}v(u) + \bs{b}\right) du + \bs{\Sigma} d\bs{B}(u),\quad u\in [t,T], \\
\bs{X}(t)=\bs{x}, \qquad \bs{x} \in \mathbb{R}^{n+1}.
\end{cases}
\end{equation}
The set of admissible controls $\mathcal{V}_{\text{ad}}$ is defined by
\begin{equation*}
\mathcal{V}_{\text{ad}}(t,\bs{x}) := \left\{ v:[t,T]\times \Omega \rightarrow \mathbb{R}\,|\,v\in \mathcal{H}_{\mathbb{F}^{t}}^{2}(t,T;\mathbb{R})\right\},
\end{equation*}
where $\mathcal{H}^{2}_{\mathbb{F}^{t}} (t,T; \mathbb{R})$ denotes the set of $\mathbb{F}^{t}$-progressively measurable $\mathbb{R}$-valued processes $\left\{H(u)\right\}_{u \in [t,T]}$ such that $\mathbb{E}\left[\int_{t}^{T}\left|H(u)\right|^{2}du \right] < +\infty$. Observe that, for any $v(\cdot) \in \mathcal{V}_{\text{ad}}(t,\bs{x})$, the above state equation admits a unique (strong) solution $\bs{X}(\cdot) := \bs{X} (\cdot; t,\bs{x},v(\cdot))$.


\subsection{The objective functional}

Given the initial data $(t,\bs{x})\in
[0,T]\times \mathbb{R}^{n+1}$, the investor's objective functional $J$ reads as
\begin{equation*}
J(t,\bs{x};v(\cdot )):=\mathbb{E}\left[ \Pi^{t}(T) -\beta X^{2}(T) -\lambda \QV{\Pi^{t}(T)}\right]
\end{equation*}
where
\begin{equation*}
\Pi^{t}(T):=\Pi^{0}(T)-\Pi^{0}(t),
\end{equation*}
namely, $\Pi^{t}(T)$ denotes the P\&L earned over the time interval $[t,T] $. Therefore, taking into account \eqref{eq:quadratic-variation}, \eqref{eq:P&L-QV-1},
\eqref{eq:parameter-P&L-QV}, and temporarily ignoring the constant term $\gamma m^{2}$, the objective functional $J$ can be written as
\begin{equation*}
J(t,\bs{x};v(\cdot )) := \Eof{\bX'(T) \bG \bX(T) + \int_{t}^{T} \left(2v(u) \bk' \bX(u) - \widetilde\eta v^{2}(u) -\psi X^{2}(u) + \mu X(u) \right) du},
\end{equation*}
where
\begin{eqnarray*}
\bG := \left[\begin{array}{cccc}
0 & \cdots & 0 & -\dfrac{\phi \nu_1}{2} \\[3pt]
0 & \cdots & 0 & -\dfrac{\phi \nu_2}{2} \\[3pt]
\vdots & \vdots & \vdots & \vdots \\[3pt]
0 & \cdots & 0 & -\dfrac{\phi \nu_n}{2} \\[3pt]
-\dfrac{\phi \nu_1}{2} & \cdots & -\dfrac{\phi \nu_n}{2} & \dfrac{\gamma}{2} - \beta
\end{array}\right], \quad
\bk := \frac{1}{2}
\left[\begin{array}{c}
-\phi\nu_{1} \\[6pt]
\vdots \\[6pt]
-\phi\nu_{n} \\[6pt]
-\widetilde\xi
\end{array}
\right].
\end{eqnarray*}
Our goal is thus to find the solution to the following linear-quadratic (LQ) stochastic control problem
\begin{problem} \label{pb:SLQ}
For any initial data $(t,\bs{x}) \in [0,T] \times \mathbb{R}^{n+1}$, maximize $J(t,\bs{x}; v(\cdot))$  subject to the state equation \eqref{eq:SDE-m0} over $v(\cdot) \in \mathcal{V}_{\textrm{ad}}(t,\bs{x})$.
\end{problem}
The optimal solution to the control problem, i.e., the value function $W$, is defined as
\begin{equation}\label{eq:value-function}
\begin{cases}
\displaystyle{W(t,\bs{x}):=\sup_{v(\cdot )\in \mathcal{V}_{\textsl{ad}}(t,x)} J(t,\bs{x};v(\cdot )),\quad \forall (t,\bs{x})\in [0,T]\times \mathbb{R}^{n+1}} \\[12pt]
W(T,\bs{x}):= \bs x'\bG \bs x, \quad \forall \bs{x}\in \mathbb{R}^{n+1},
\end{cases}
\end{equation}


\subsection{The HJB equation}
By standard arguments in stochastic control theory, the value function $W$ satisfies a Hamilton-Jacobi-Bellman (HJB) equation. In our framework, the HJB equation in $[0, T) \times \mathbb{R}^{n+1}$ with terminal boundary condition is given by
\begin{equation*}
\begin{cases}
\displaystyle \frac{\partial}{\partial t}w(t,\bs{x}) + \mathcal{H} \left(\bs{x}, \bs{D} w(t,\bs{x}), \bs{D}^{2}w(t,\bs{x}) \right) = 0, \quad (t,\bs{x}) \in [0,T) \times \mathbb{R}^{n+1}, \\[12pt]
w(T,\bs{x}) = \bs x' \bG \bs x, \quad \forall \bs{x} \in \mathbb{R}^{n+1},
\end{cases}
\end{equation*}
where $\bs{D} w$ and $\bs{D}^{2}w$ denote respectively the gradient and the Hessian matrix of $w$ respect to state variable $\bs{x}$, and $\mathcal{H}$ is the Hamiltonian given by
\begin{equation*}
\mathcal{H} \left(\bs{x}, \bs{D} w(t,\bs{x}), \bs{D}^{2}w(t,\bs{x}) \right) := \sup_{v \in \mathbb{R}} \mathcal{H}_{cv} \left(\bs{x}, \bs{D} w(t,\bs{x}), \bs{D}^{2}w(t,\bs{x}); v\right) , \quad (t,\bs{x}) \in [0,T) \times \mathbb{R}^{n+1},
\end{equation*}
with $\mathcal{H}_{cv}$ representing the Hamiltonian current-value defined as
\begin{equation} \label{eq:Hcv}
\mathcal{H}_{cv} \left(\bs{x}, \bs{p}, \bs{P}; v\right) := 2v \bk'\bs x - \widetilde{\eta} v^{2} - \psi x^2 + \mu x + \bs{p}'\left(\bs{A}\bs{x} + \bs{a} v + \bs{b}\right) +\frac{1}{2}\tr\left[ \bs{\Sigma}\bs{\Sigma}' \bs{P} \right]
\end{equation}
for $(\bs{x}, \bs{p},\bs{P}) \in \mathbb{R}^{n+1} \times \mathbb{R}^{n+1} \times \mathcal{S}^{n+1}, v \in \mathbb{R}$, where $\mathcal{S}^{n+1}$ is the set of $(n+1) \times (n+1)$ symmetric matrices.

Given $(\bs{x}, \bs{p},\bs{P}) \in \mathbb{R}^{n+1} \times \mathbb{R}^{n+1} \times \mathcal{S}^{n+1}$, the function $v \rightarrow \mathcal{H}_{cv} \left(\bs{x}, \bs{p}, \bs{P}; v\right)$ has a unique maximum point over $\mathbb{R}$ given by
\begin{equation}\label{eq:maxpoint}
v^{\star}(\bs{x},\bs{p}) = \frac{2 \bk'\bs x + \bs{a}' \bs p}{2\widetilde{\eta}}.
\end{equation}
The HJB equation associated to the stochastic control problem \eqref{pb:SLQ} reduces to
\begin{subequations}\label{eq:HJB}
\begin{multline}\label{eq:HJB-matrix-form}
\frac{\partial}{\partial t}w(t,\bs{x}) + \frac{1}{4\widetilde{\eta}} \left(2\bk'\bs x + \bs a' \bs{D} w(t,\bs{x})\right)^{2} \\
- \psi x^2 + \mu x + \left(\bs{A}\bs{x} + \bs{b}\right)'\bs{D} w(t,\bs{x}) + \frac{1}{2}\tr\left[ \bs{\Sigma}\bs{\Sigma}' \bs{D}^{2} w(t,\bs{x}) \right] = 0, \quad (t,\bs{x}) \in [0,T) \times \mathbb{R}^{n+1},
\end{multline}
with terminal condition
\begin{equation}\label{eq:HJB-tc}
w(T,\bs{x}) = \bs x'\bs{G} \bs{x}, \quad \forall \bs{x} \in \mathbb{R}^{n+1}.
\end{equation}
\end{subequations}

The value function $W$ defined in \eqref{eq:value-function} may be characterized as the unique solution to \eqref{eq:HJB}.

\begin{definition}[Classical solution]\label{def:class-sol}
A function $w$ is called a classical solution of \eqref{eq:HJB} if
\begin{enumerate}[(i)]
\item $w\in C^{1,2}([0,T]\times\mathbb{R}^{n+1};\mathbb{R})$
\item $w$ satisfies pointwise in classical sense \eqref{eq:HJB-matrix-form}
    (the derivatives with respect to time variable at $t=0$ and $t=T$ have to be intended respectively as right and left derivative),
\item $w$ satisfies the boundary condition \eqref{eq:HJB-tc}. \hfill$\square$
\end{enumerate}
\end{definition}

\begin{lemma}[Solution to HJB equation]\label{lemma:expl-sol-HJB}
Let $\beta > \dfrac{\gamma}{2}$. Define
\begin{equation}\label{eq:expl-sol-HJB}
w(t, \bs{x}) = \bs{x}' \bs{R}(t)\bs{x} + \bs{r}'(t) \bs{x} + \varphi(t), \quad \forall (t,\bs{x}) \in [0,T] \times\mathbb{R}^{n+1},
\end{equation}
where $\bs{R} \in C([t,T]; \mathcal{S}^{n+1})$, $\bs{r} \in C([t,T]; \mathbb{R}^{n+1})$, and $\varphi \in C([t,T]; \mathbb{R})$ are the unique solution to the following terminal value problem (TVP), for some $t \in [0,T]$,
\begin{subnumcases}{\label{eq:Riccati-system}}
\dot \bR - \psi \be_{n+1} \be'_{n+1} + \frac1{\widetilde\eta} \left\{ \bR \ba \ba'\bR + (\widetilde\eta \bA + \ba \bk')' \bR + \bR (\widetilde\eta \bA + \ba \bk') + \bk\bk' \right\} = 0 \label{eq:Riccati-n} \\
\dot \br + \bA'\br + 2 \bR \bb + \mu\be + \frac1{\widetilde\eta}\left(\bR \ba \ba' + \bk\ba'\right) \br = 0 \label{eq:linear-term-n} \\
\dot \varphi + \tr(\bs\Sigma\bs\Sigma'\bR) + \bb'\br + \frac1{4\widetilde\eta} \br'\ba\ba'\br = 0 \label{eq:constant-term-n}
\end{subnumcases}
with terminal condition
\begin{equation}\label{eq:Riccati-TC}
\bR(T) = \bG, \qquad \br(T) = \boldsymbol 0, \qquad \varphi(T) = 0,
\end{equation}
where for notational simplicity we suppressed the dependence on $t$ for the functions $\bR$, $\br$, and $\varphi$ unless necessary, and $\be_{n+1}$ denotes the unit vector $\be'_{n+1} = \left[0 \; \cdots 0 \; 1 \right] \in \R^{n+1}$. Then the quadratic function $w$ in the state variable $\bs{x}$ is a classical solution to the HJB equation \eqref{eq:HJB}.
\end{lemma}

\begin{proof}
Substituting \eqref{eq:expl-sol-HJB} into \eqref{eq:Hcv} with $\bs p = \bs D w(t, \bs x) = 2\bR\bs x + \br$, $\bs P = \bs D^2 w(t, \bs x) = 2\bR$ and completing the square for $v$, the Hamiltonian current value $\mathcal{H}_{cv}$ reads as
\begin{equation}\label{eq:Hcv-squared}
\begin{aligned}
&\mathcal{H}_{cv} \left(\bs{x}, \bs{D} w(t,\bs{x}), \bs{D}^{2} w(t,\bs{x}); v\right) \\
&= -\widetilde{\eta} \left[v - \frac{1}{2\teta}(2\bs k + 2\bs R\bs a)' \bs x - \frac{1}{2\widetilde{\eta}} \bs a'\bs r\right]^{2} +\frac1{4\teta}\left\{(2\bs k + 2\bs R\bs a)' \bs x + \bs a'\bs r \right\}^2 \\
&\quad -\psi x^2 + \mu x + (2\bs R \bs x+ \bs r)'(\bs A \bs x + \bs b) + \tr\left[ \bs{\Sigma}\bs{\Sigma}'\bs{R}\right].
\end{aligned}
\end{equation}
Since $\widetilde{\eta}>0$, from \eqref{eq:Hcv-squared} we clearly see that the function $v \rightarrow \mathcal{H}_{cv} \left(\bs{x}, \bs{D} w, \bs{D}^{2} w; v\right) $ has a unique maximum point $v^\star$ over $\mathbb{R}$ given by
\begin{equation*}
v^{\star} = \frac{2\left\{\bk + \bR \ba\right\}' \bs{x} + \ba' \bs r}{2\widetilde{\eta}}
, \qquad (t,\bs{x}) \in[0,T] \times \mathbb{R}^{n+1},
\end{equation*}
and the HJB equation reads as
\begin{equation*}
\bs x' \dot{\bs{R}}\bs{x} + \bs x' \dot{\bs{r}} + \dot{\varphi} = \frac{1}{4\teta}\left\{(2\bs k + 2\bs R\bs a)' \bs x + \bs a'\bs r \right\}^2 -\psi x^2 + \mu x + (2\bs R \bs x+ \bs r)'(\bs A \bs x + \bs b) + \tr\left[ \bs{\Sigma}\bs{\Sigma}' \bs{R}\right]
\end{equation*}
with terminal condition
\begin{equation*}
w(T,\bs{x}) = \bs x' \bs{G} \bs{x}, \qquad \forall \bs{x} \in \mathbb{R}^{n+1}.
\end{equation*}
Finally, by comparing the coefficients, we obtain the Riccati equation \eqref{eq:Riccati-n}, the linear term \eqref{eq:linear-term-n}, and the constant term \eqref{eq:constant-term-n}, respectively, with
\begin{equation*}
\bs{R}(T) = \bs{G}, \qquad \bs{r}(T) = \textbf{0}, \qquad \varphi(T) = 0.
\end{equation*}
\end{proof}
We remark that the system of ODEs \eqref{eq:Riccati-n} satisfied by $\bR$ is a matrix Riccati differential equation whereas the ODEs for $\br$ and $\varphi$ are linear. Thus, the existence and uniqueness for $\br$ and $\varphi$ are straightforward so long as we have those for $\bR$. The flow of Riccati equation \eqref{eq:Riccati-n} can be linearized by doubling the dimension of the problem. This is due to the fact that a Riccati ODE solution belongs to a quotient manifold (see \textcite{GraTeb} for further details). The explicit linearization procedure and the closed form solution to \eqref{eq:Riccati-n} follows from \textcite{DaFGraTeb}. It is interesting to notice however that the explicit computation is made non-trivial by the fact the computation of matrix exponentials is complicated by the presence of degenerate matrices. Computation of $\bR$ is linearized thanks to the following lemma.
\begin{lemma}(Solution to the matrix Riccati equation) \\
The solution to the matrix Riccati equation \eqref{eq:Riccati-n} for $\bR$ solves the linear system
\begin{equation*}
\bR N(t)= M(t), \quad t\in [0,T]
\end{equation*}
where the $(n+1)\times(n+1)$ matrices $M$ and  $N$ satisfy the following system of linear ODEs
\begin{equation}
\frac{d}{dt} \left[\begin{array}{c} M(t) \\ N(t) \end{array}\right] =
\left[\begin{array}{cc} -\left(\bA + \dfrac{1}{\widetilde\eta}\ba\bk'\right)' & \psi\be\be' - \dfrac{1}{\widetilde\eta}\bk\bk' \\
\dfrac{1}{\widetilde\eta}\ba\ba' & \bA + \dfrac{1}{\widetilde\eta}\ba \bk'
\end{array}\right] \,
\left[\begin{array}{c} M(t) \\ N(t) \end{array}\right], \quad t\in [0,T] \label{eq:MN-odes}
\end{equation}
with terminal conditions $M(T) = \bG$ and $N(T) = \bs I$.
Moreover, the solution to the linear system can be written as
\begin{eqnarray*}
\left[\begin{array}{c} M(t) \\ N(t) \end{array}\right] =e^{-(T-t)\Psi} \,
\left[\begin{array}{c} \bG \\ \bs I \end{array}\right],
\end{eqnarray*}
where
\begin{eqnarray*}
\Psi := \left[\begin{array}{cc} -\left(\bA + \dfrac{1}{\widetilde\eta}\ba\bk'\right)' & \psi\be\be' - \dfrac{1}{\widetilde\eta}\bk\bk' \\
\dfrac{1}{\widetilde\eta}\ba\ba' & \bA + \dfrac{1}{\widetilde\eta}\ba \bk'
\end{array}\right].
\end{eqnarray*}
and $\bs I$ is an $(n+1)\times (n+1)$ identity matrix.
\end{lemma}
\begin{proof}
By straightforward substitution one can show that if $M$ and $N$ satisfy the system of ODEs \eqref{eq:MN-odes}, then a solution to the linear system $\bR N(t)= M(t)$ solves \eqref{eq:Riccati-n}.
\end{proof}

We present closed form expressions for optimal trading rates $v^\star$ in this case, in the next Corollary \ref{cor:cor1} and Corollary \ref{cor:cor2}.


\subsection{The verification theorem and the optimal feedback policy} \label{sec:verification}

The aim of this section is to prove a verification theorem stating that the function $w$ defined in \eqref{eq:expl-sol-HJB} is actually the value function and giving an optimal feedback strategy for the stochastic LQ problem \ref{pb:SLQ}.

Let $(t,\bs{x})\in [0,T]\times \mathbb{R}^{n+1}$, take into account \eqref{eq:maxpoint} and Lemma \ref{lemma:expl-sol-HJB}. The feedback map $G$ reads as
\begin{equation*}
\left(t,\bs{x}\right) \mapsto G(t,\bs{x}):= \frac{2\bk'\bs x + \ba' \bs{D} w(t,\bs x)}{2\widetilde{\eta}} .
\end{equation*}
By virtue of Lemma \ref{lemma:expl-sol-HJB} we have
\begin{equation*}
\bs{D} w(t,\bs{x}) = 2\bs{R}(t)\bs{x} + \bs{r} (t), \qquad \forall (t,\bs{x}) \in [0,T]\times \mathbb{R}^{n+1},
\end{equation*}
the above feedback map becomes
\begin{equation}\label{eq:feedback}
G(t,\bs{x})= \frac{2(\bk + \bR(t) \ba)'\bs{x} + \ba' \br(t)}{2\widetilde{\eta}}, \qquad (t,\bs{x}) \in[0,T] \times \mathbb{R}^{n+1}.
\end{equation}

The corresponding closed loop equation is
\begin{equation}\label{eq:CLE}
\begin{cases}
d\bs{X}(u) = \left[\left\{\bs{A} + \dfrac{1}{\widetilde{\eta}} \left( \bs{a} \bk' + \ba \ba' \bR(u) \right)\right\}\bs{X}(u) +\dfrac{1}{2\widetilde{\eta}} \, \bs{a} \bs a' \bs{r}(u) + \bs{b}\right] du + \bs{\Sigma} d\bs{B}(u),\quad u\in [t,T], \\
\bs{X}(t)=\bs{x}, \quad \bs{x} \in \mathbb{R}^{n+1}.
\end{cases}
\end{equation}

\begin{lemma}[Solution to closed loop equation]\label{lemma:CLE}
For every $(t,\bs{x}) \in[0,T] \times \mathbb{R}^{n+1}$ there exists a unique $\mathbb{F}^{t}$-progressively measurable process $\bs{X}_{G}(\cdot ; t,\bs{x}) \in L^{2}\left(\Omega \times[t,T]; \mathbb{R}^{n+1}\right)$ solution to \eqref{eq:CLE}.
\end{lemma}

\begin{proof}
The proof of the existence and uniqueness of $\bs{X}_{G}(\cdot;t,\bs{x})$ is due to the Lipschitz continuity of the map $G$ and it is a rather standard result (see, e.g., \textcite[Chapter~5, Theorem~2.5, p.~287, and Theorem~2.9, p.~289]{KarShr}). \end{proof}

By applying standard arguments we obtain the following result.

\begin{theorem}[Verification theorem and optimal feedback]\label{th:VT}
Let $\beta >\dfrac{\gamma }{2}$. Then the function $w$ given in \eqref{eq:expl-sol-HJB} is the value function $W$ defined in \eqref{eq:value-function}, namely,
\begin{equation*}
W(t,\bs{x})=w(t,\bs{x}), \qquad \forall (t,\bs{x})\in [0,T]\times\mathbb{R}^{n+1}.
\end{equation*}
Furthermore, $v^{\star}(\cdot)\in\Vad(t,\bs{x})$ is optimal for the initial point $(t,\bs{x})\in [0,T]\times\mathbb{R}^{n+1}$  if and only if
\begin{equation}\label{eq:optimal-feedback}
v^{\star}(u) = G(u,\bs{X}_{G}(u;t,\bs{x}), \qquad u\in[t,T],
\end{equation}
where $G$ is given by \eqref{eq:feedback} and $\bs{X}_{G}(\cdot;t,\bs{x}))$ is the unique solution to the closed loop equation \eqref{eq:CLE}, i.e.,
\begin{equation}\label{eq:optimal-strategy}
v^{\star}(u) = \frac{2(\bk + \bR(u) \ba)'\bs{X}_{G}(u ; t,\bs{x}) + \ba' \br(u)}{2\widetilde{\eta}}, \qquad u\in[t,T].
\end{equation}
In particular, the above feedback strategy
$v^{\star}$ is the unique optimal strategy.
\end{theorem}

\begin{proof}
By Lemma \ref{lemma:expl-sol-HJB}, the function $w$ defined in \eqref{eq:expl-sol-HJB} is a classical solution to the HJB equation \eqref{eq:HJB}. Let $(t,\bs{x})\in [0,T]\times \mathbb{R}^{n+1}$, $v(\cdot)\in \Vad(t,\bs{x})$, and apply the Dynkin formula to the corresponding state trajectory $\bs{X}(\cdot):= \bs{X} \left(\cdot; t,\bs{x},v(\cdot)\right)$ with the function $w$. We obtain
\begin{align*}
&w(T,\bs{X}(T))- w(t,\bs{x}) \\
&= \mathbb{E} \left[\int_{t}^{T}
\left\{ \frac{\partial}{\partial t} w(u,\bs{X}(u)) + (\bs{A}\bs{X}(u) + \bs{a} v(u) + \bs{b})'\bs{D} w(u,\bs{X}(u)) +\frac{1}{2}\tr\left[\bs{\Sigma}\bs{\Sigma}' \bs{D}^{2} w(u,\bs{X}(u)) \right]du \right\} \right],
\end{align*}
i.e.,
\begin{align*}
&w(t,\bs{x}) = \mathbb{E} \Bigg[ \bs X(T)' \bs G \bs X(T) \\
&- \int_{t}^{T}
\left\{ \frac{\partial}{\partial t} w(u,\bs{X}(u)) + (\bs{A}\bs{X}(u) + \bs{a} v(u) + \bs{b})'\bs{D} w(u,\bs{X}(u)) +\frac{1}{2}\tr\left[\bs{\Sigma}\bs{\Sigma}' \bs{D}^{2} w(u,\bs{X}(u)) \right]du \right\} \Bigg].
\end{align*}
Recalling that $w$ solves the original HJB equation \eqref{eq:HJB}, we may write the fundamental identity
\begin{multline*}
w(t,\bs{x})= J(t,\bs{x};v(\cdot)) + \mathbb{E} \left[\int_{t}^{T} \left( \mathcal{H} \left(\bs{X}(u), \bs{D} w(u,\bs{X}(u)), \bs{D}^{2} w(u,\bs{X}(u))\right. \right.\right.\\
\left.- \mathcal{H}_{cv} \left(\bs{X}(u), \bs{D} w(u,\bs{X}(u)), \bs{D}^{2} w(u,\bs{X}(u)); v(u) \right)\right)du \Bigg],
\end{multline*}
obtaining
\begin{equation}\label{eq:w-geq-J}
w(t,\bs{x}) \geq  J(t,\bs{x};v(\cdot)).
\end{equation}
As the above inequality holds for every $v(\cdot) \in \Vad(t,\bs{x})$ and $\mathcal{H}(\cdot) \geq \mathcal{H}_{cv}(\cdot)$ for every $v \in \mathbb{R}$, thus $w \geq W$.

Now, consider $\bs{X}_{G}(\cdot) := \bs{X}(\cdot;t,\bs{x},v^{\star}(\cdot))$ and apply the fundamental identity to $\bs{X}_{G}(\cdot)$ with function $w$. Taking into account Lemma \ref{lemma:expl-sol-HJB} and \eqref{eq:feedback} we see that the feedback map maximizes at any time $t\in[0,T]$ the Hamiltonian current value. Thus, in this case we have $w(t,\bs{x}) = J(t,\bs{x};v^{\star}(\cdot))$, which shows that
\begin{equation*}
w(t,\bs{x}) = W(t,\bs{x}) = J(t,\bs{x}; v^{\star}(\cdot)).
\end{equation*}

By the uniqueness of the solution to the closed loop equation \eqref{eq:CLE}  stated in Lemma \ref{lemma:CLE} and the Lipschitz continuity of $G$, the feedback strategy $v^{\star}(\cdot)$ is admissible, that is, $v^{\star}(\cdot)\in\Vad(t,\bs{x})$. Furthermore, an optimal strategy must satisfy \eqref{eq:optimal-feedback} since $W=w$ and \eqref{eq:w-geq-J} holds. Finally, the uniqueness  of the optimal strategy is a consequence of the characterization \eqref{eq:optimal-feedback} and uniqueness of solution to the closed loop equation \eqref{eq:CLE}.
\end{proof}

\subsection{Optimal trading strategy and closed form solutions to the matrix Riccati equation}

Let $\bX_{G}(\cdot; t, \bs{x}) := \left[\bs{Q}_{G}(\cdot), X_{G}(\cdot)\right]'$ the solution to the closed loop equation \eqref{eq:CLE}. The following two corollaries hold.

\begin{corollary}[$\bs\bq_1 = \bs 0$ and $\lambda > 0$] \label{cor:cor1}
Let $\bs\bq_1 = \bs 0$. In this case, $\ba = -\bs{e}_{n+1}$. Assume $\left(\frac{\gamma + \widetilde{\xi}}{2} - \beta\right)^2 > \psi\teta$ and $\beta - \frac{\gamma + \widetilde{\xi}}{2} > 0$. Define the constant $\talpha$ by
\begin{eqnarray}
\talpha := \sqrt{\dfrac\teta\psi} \sinh^{-1}\left\{\dfrac{\sqrt{\psi\teta}}{\sqrt{\left|\left(\frac{\gamma+\widetilde{\xi}}2 - \beta\right)^2 - \psi\teta\right|}} \right\}. \label{eq:tilde-alpha}
\end{eqnarray}
Thus, we have
\begin{eqnarray*}
&& \sinh\left(\sqrt{\dfrac\psi\teta}\talpha\right) = \dfrac{\sqrt{\psi\teta}}{\sqrt{\left| \left(\frac{\gamma+\widetilde{\xi}}2 - \beta\right)^2 - \psi\teta \right|}} \qquad \text{and} \qquad
\cosh\left(\sqrt{\dfrac\psi\teta}\talpha\right) = \dfrac{\beta -\dfrac{\gamma+\widetilde{\xi}}2}{\sqrt{\left|
\left(\frac{\gamma+\widetilde{\xi}}2 - \beta\right)^2 - \psi\teta \right|}}.
\end{eqnarray*}
Let $\zeta := \sqrt{\dfrac\psi\teta}$.
The optimal trading rate $v^\star$ in this case is given by the feedback form
\begin{eqnarray*}
v^\star(u) &=& \zeta \coth\left(\zeta\left\{T - u + \talpha\right\}\right) X_{G}(u) - \dfrac{\phi}{2\teta}\bs\nu'\bs{Q}_{G}(u) + \dfrac{\phi}{2\teta} \bs\nu' \left\{\bs\Lambda^{u}(T) \bs{Q}_{G}(u) + \bs\Lambda^{u}_{0}(T) \bs\bq_0\right\} \\
\quad &&+\dfrac\mu{2\sqrt{\psi\teta}} \left\{ \dfrac{\cosh(\zeta\talpha)}{\sinh(\zeta(T - u + \talpha))} - \coth(\zeta(T - u + \talpha)) \right\},
\end{eqnarray*}
where $\bs\Lambda^{u}(T)$ and $\bs\Lambda^{u}_{0}(T)$ are time dependent $n\times n$ matrices defined by
\begin{multline}\label{eq:Pi-t}
\bs\Lambda^{u}(T) := \dfrac{\sinh(\zeta\talpha)}{\sinh(\zeta(T - u + \talpha))} e^{(u-T)\bs\Theta} \left[ \bs I - \left(\bs I - \dfrac1{\zeta^2}\bs\Theta^2\right)^{-1} + \dfrac1{\zeta} \left(\bs I - \dfrac1{\zeta^2}\bs\Theta^2\right)^{-1} \bs\Theta \coth(\zeta\talpha) \right] \\
+ \left(\bs I - \dfrac1{\zeta^2}\bs\Theta^2\right)^{-1} - \dfrac1{\zeta} \left(\bs I - \dfrac1{\zeta^2}\bs\Theta^2\right)^{-1} \bs\Theta \coth(\zeta(T - u + \talpha))
\end{multline}
and
\begin{multline}\label{eq:Pi0-t}
\bs\Lambda^{u}_{0}(T) \\
:= \dfrac{\sinh(\zeta\talpha)}{\sinh(\zeta(T - u + \talpha))} \left[ \bs I - e^{(u-T)\bs\Theta} \right] \left[ \bs I - \left(\bs I - \dfrac1{\zeta^2}\bs\Theta^2\right)^{-1} + \dfrac1{\zeta} \left(\bs I - \dfrac1{\zeta^2}\bs\Theta^2\right)^{-1} \bs\Theta \coth(\zeta\talpha) \right] \\
+ \dfrac1\zeta \left(\bs I - \dfrac1{\zeta^2}\bs\Theta^2\right)^{-1} \bs\Theta \left[-\dfrac{\cosh(\zeta\talpha)}{\sinh(\zeta(T - u + \talpha))} + \coth\left\{\zeta(T - u + \talpha)\right\} \right] \\
-\dfrac1{\zeta^2} \left(\bs I -\dfrac1{\zeta^2}\bs\Theta^2\right)^{-1} \bs\Theta^2 \left[ 1 -\dfrac{\sinh(\zeta\talpha)}{\sinh(\zeta(T - u + \talpha))} \right]
\end{multline}
\end{corollary}
\begin{proof}
The proof is lengthy and tedious. We postpone it to Section \ref{app:proof-cor1} in Appendix.
\end{proof}

\begin{remark}
In the above corollary we assumed the conditions $\left(\frac{\gamma + \widetilde{\xi}}{2} - \beta\right)^2 > \psi\teta$ and $\beta - \frac{\gamma + \widetilde{\xi}}{2} > 0$ that are sufficient for our discussion but not necessary. All the other cases can be also discussed with minor modifications.
\end{remark}

Note that, when $\phi = 0$, i.e., price impact not taking into account the inventory cost component, the optimal trading rate $v^\star$ given above recovers the optimal trading rate obtained in \textcite[p.~62,~eq.~(4.7)]{QF} which reads apparently, in the current notations,
\begin{equation*}
v^{\star}(u) = \zeta \coth\left(\zeta\left\{T - u + \talpha\right\}\right) X_{G}(u) + \frac\mu{2\sqrt{\psi\teta}} \left\{ \frac{\cosh(\zeta\talpha)}{\sinh(\zeta(T - u + \talpha))} - \coth(\zeta(T - u + \talpha)) \right\}.
\end{equation*}

Optimal trading strategy for a risk neutral trader, i.e., $\lambda = 0$, when $\bs\bq_1 = \bs 0$ can be further reduced and the final expression becomes much neater. We summarize the result in the following corollary whose proof will be omitted since it can be obtained by repeating the procedure as in Corollary \ref{cor:cor1} with $\lambda = 0$.
\begin{corollary}[$\bs\bq_1 = \bs 0$ and $\lambda = 0$]\label{cor:cor2}
Let $\lambda \equiv 0$ and $\bs\bq_1 \equiv \bs 0$. In this case, we have $\widetilde\eta = \eta$, $\psi = \widetilde{\xi} = 0$, and $\ba = -\bs{e}_{n+1}$. The optimal trading rate is given in closed form by
\begin{align*}
v^{\star}(u) &= \frac\phi{2\eta}(\bs\nu'\bs\bq_0 - \bs\nu'\bs{Q}_{G}(u)) + \frac\phi{2\eta} \frac{\bs\nu'}{T - u + \alpha} \left[\alpha e^{-(T - u)\bs\Theta} + \left(e^{-(T - u)\bs\Theta} - I\right) \bs\Theta^{-1} \right] (\bs{Q}_{G}(u) - \bs\bq_0) \\
&\quad+ \frac {X_{G}(u)}{T - u + \alpha} - \frac\mu{4\eta} \left( T - u + \alpha - \frac{\alpha^2}{T - u + \alpha}\right).
\end{align*}
\end{corollary}

Note that the optimal trading rate $v^\star$ depends on the remaining shares to be liquidated $x$ and the discrepancy between the current inventory $\bs q$ and its long term mean $\bs\bq_0$. Thus, the trader is suggested to take into account the traded volume while liquidating his position optimally. We remark that apparently when $\phi = 0$, i.e., price impact disregarding the inventory cost component, then the optimal trading rate $v^\star$ reduces to
\begin{equation*}
v^\star(u) = \frac {X_{G}(u)}{T - u + \alpha} - \frac\mu{4\eta} \left( T - u + \alpha - \frac{\alpha^2}{T - u + \alpha}\right)
\end{equation*}
which recovers the optimal trading rate in \textcite[p.~58,~eq.~(3.3)]{QF} .

\section{Numerical examples} \label{sec:numerical}
In order to illustrate the performance of the optimal liquidation strategy obtained in Theorem \ref{th:VT} and gain some economic insight, in this section we run a number of numerical tests to assess the marginal improvement of the current approach with respect to known, benchmark execution strategies.

The investor's target is assumed to be the liquidation within one day of the amount of $x_0 = 200,000$ shares of a certain stock. Parameters for price impact are selected so as to be in line with those in \textcite{AlmChr} and \textcite{QF,JORS}: $\gamma$ and $\eta$ are set as $\gamma=2.5\times 10^{-7}$ and $\eta=2.5\times 10^{-6}$. We set the order-fill-uncertainty parameters as $m = x_0\times10\%=20,000$. This is equivalent to say that the execution risk may generate on average a deviation from the submission path of $10\%$. The initial stock price level, irrelevant for the implementations of the strategies under consideration, is assumed $S_0=50$, price volatility is set to $\sigma_S = 0.5$ and for simplicity expected annual return for the stock is set to zero.

The spectrum of market makers consists of 10 market makers indexed by the mean-reverting rates $\theta_i = i$ for $i = 1, 2, \cdots, 10$. The weight $\nu_i$ is chosen as a discretized gamma distribution with degrees of freedom 3. Specifically,
\begin{equation*}
\nu_i = \frac{\Gamma(i, 3)}{\sum_{n=1}^{10}\Gamma(n, 3)},
\end{equation*}
where $\Gamma(\cdot, 3)$ denotes the probability density function for gamma distribution with degrees of freedom 3. The long term means of the Ornstein-Uhlenbeck processes are assumed
\begin{eqnarray*}
&& \bq_i(t) = \frac {v(t)}{100\, \theta_i} + \frac{1}{10\, \theta_i}
\end{eqnarray*}
for $i \in \{1, 2, \cdots, 10\}$. The rationale for this choice is that the higher the market maker's mean-reverting rate, the faster the market maker is able to quickly reduce its inventories and the closer the long term mean of the inventory to zero. In fact, the market makers' objective is to carry no position overnight in average and are committed to maintain their long term expected inventory as small as possible. Recalling that the moment generating function of a gamma distribution is a power function, it is easy to verify that this choice implies a power law decay of the price impact function.
The parameters are summarized in Table \ref{tab:params}.

\bigskip

\begin{longtable}{*{4}{cccc}}
  \caption{Selected parameters in the numerical simulation.}
  \label{tab:params} \\
    \hline
    \hline
    \addlinespace
   {\bf Volatilities} & {\bf Price impact} & {\bf Block trade penalty} & \multicolumn{1}{c}{\bf Inventory} & {\bf Risk aversion} \\
    \addlinespace
	\hline
	\hline
    \addlinespace
	\begin{tabular}{l} $\begin{aligned}[t]
    &\sigma_S = 0.5 \\
    &\sigma_M = 0.1 \\
    &m = 20,000
    \end{aligned}$ \end{tabular}
	& \begin{tabular}{l} $\begin{aligned}[t]
    &\gamma = 2.5 \times 10^{-7} \\
    &\eta = 2.5 \times 10^{-6}
    \end{aligned}$ \end{tabular}
    & \begin{tabular}{l} $\beta = 100\,\eta$ \end{tabular}
    & \begin{tabular}{l} $\phi = 100\,\eta$ \end{tabular}
    & \begin{tabular}{l} $\begin{aligned}[t]
    & \lambda = 0 \\
    & \lambda = 0.001
    \end{aligned}$  \end{tabular} \\
     \addlinespace
\hline
\hline
\end{longtable}

Simulations for evaluating the objective functional \eqref{eq:P&L-QV-1} were conducted by applying the following strategy: first we focus on a comparison restricted to the performance of different strategies in a risk-neutral setup where $\lambda = 0$ and then we move on to the case where the agent is risk averse and  the risk aversion coefficient $\lambda$ is set to $0.001$. When $\lambda = 0$ we compare the optimal strategy in Theorem \ref{th:VT}, with two strategies: the one denoted TWAP obtained by setting $v(t) = \frac{x_0}T$, and a second one denoted by adapted TWAP (for risk neutral trader) which is the optimal one when $\phi=0$ and $\lambda = 0$ (the case corresponding to the analysis of \textcite{QF})
\begin{equation*}
v(t) = \frac{X(t)}{T - t + \alpha},
\end{equation*}
where $\alpha = \frac{2\eta}{2\beta - \gamma}$ and $X(t)$ denotes the remaining shares to be liquidated at time $t$.

Then we move to the more relevant case where the risk aversion is set to a finite level $\lambda = 0.001$ and we consider, in addition to the previous benchmarks also the Almgren-Chriss  one.

\begin{figure}[h!]
\caption{Expected trading trajectory during the course of execution for risk neutral trader. Optimal in red and TWAP in blue.\\} \label{fig:case2}
\includegraphics[height=6cm,width=9cm]{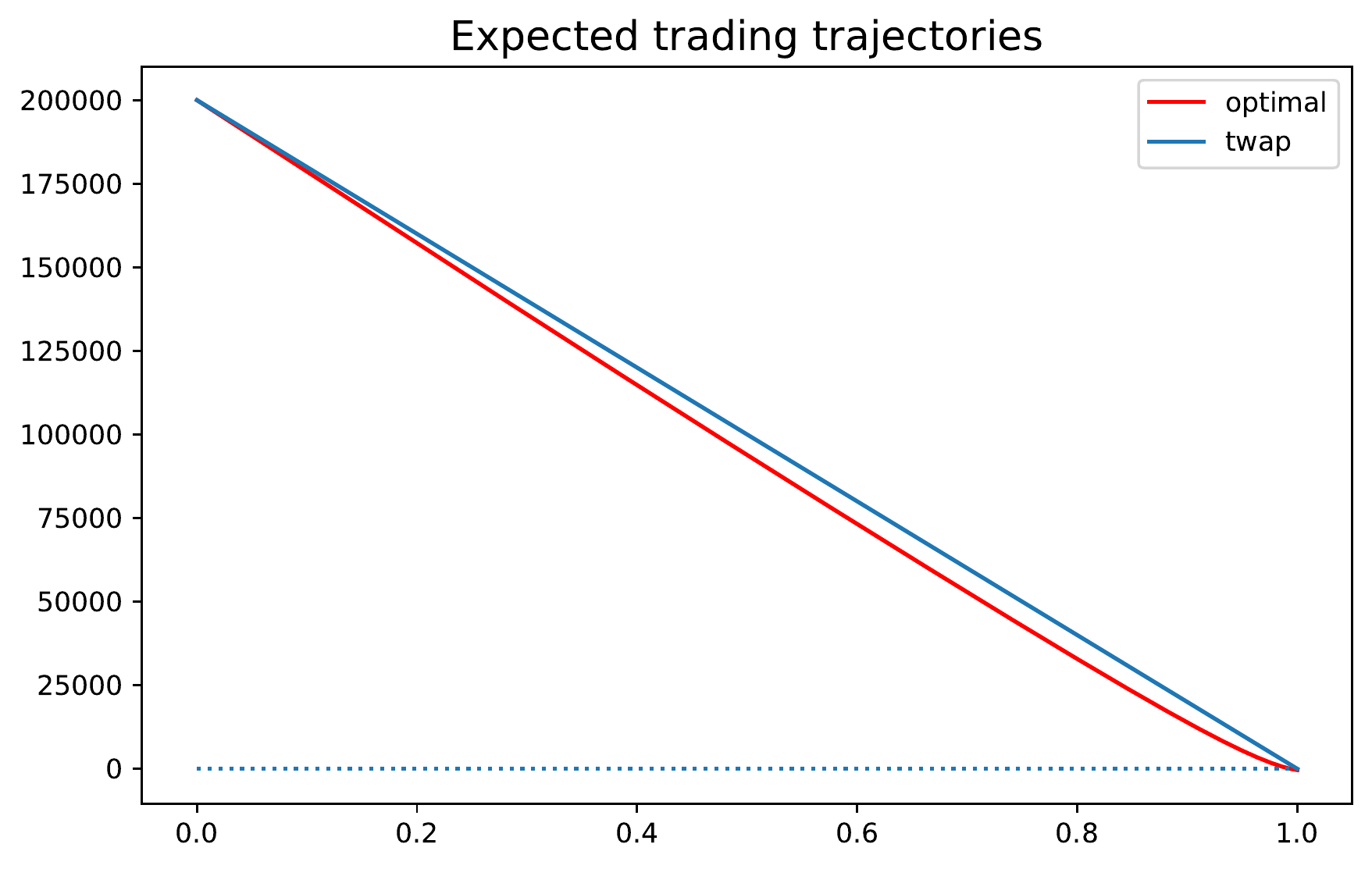}
\end{figure}

\begin{figure}[h!]
\caption{Histogram with kernel density estimate of objective functional for risk neutral trader. Optimal in green, TWAP in blue, and adapted TWAP in red.\\} \label{fig:value_hist_case2}
\includegraphics[height=6cm,width=9cm]{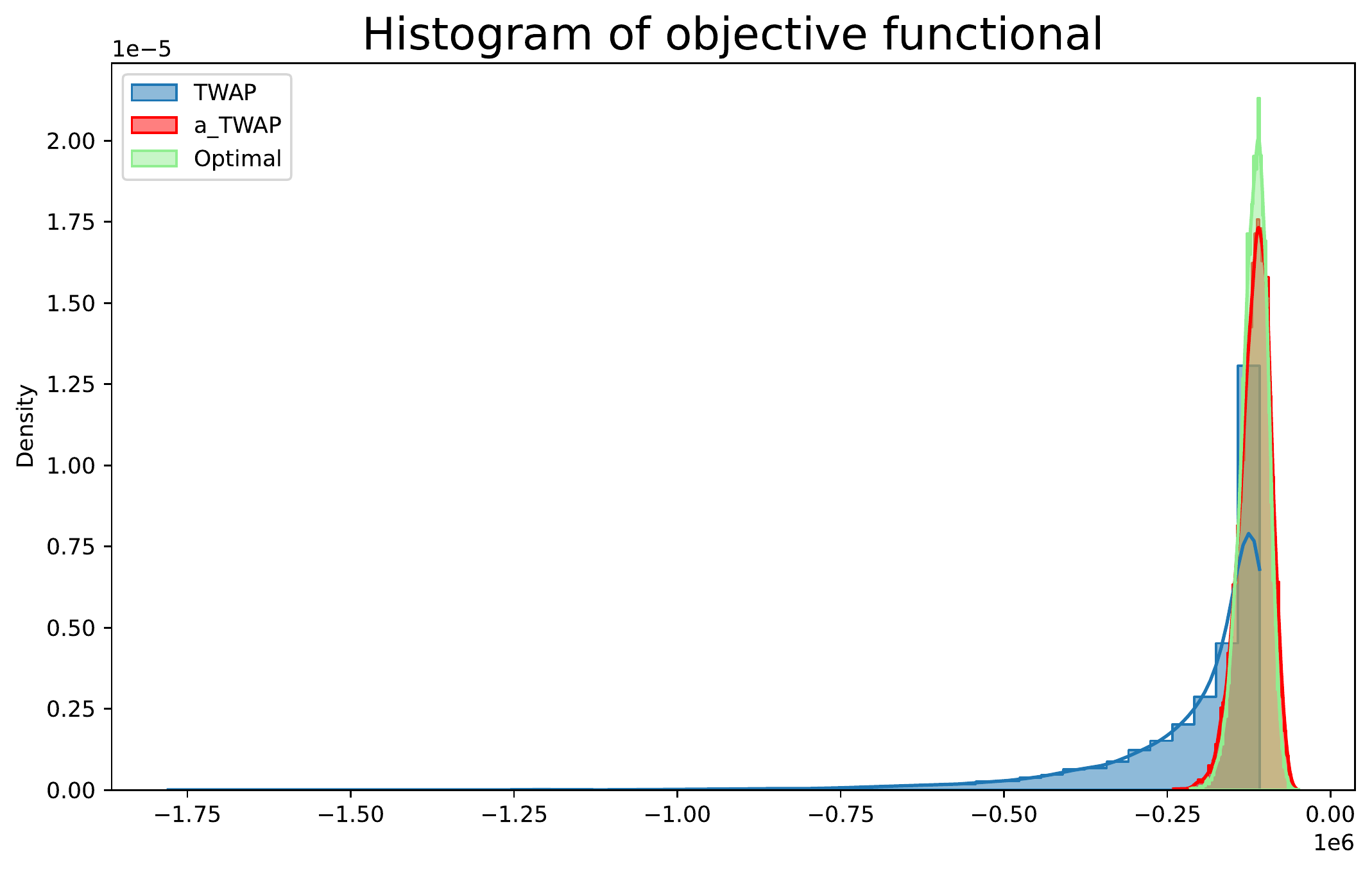}
\end{figure}

\begin{figure}[h!]
\caption{Histogram with kernel density estimate for terminal position $X_T$ to be liquidated by a block trade for risk neutral trader. Optimal in green, TWAP in blue, and adapted TWAP in red.\\} \label{fig:xT_hist_case2}
\includegraphics[height=6cm,width=9cm]{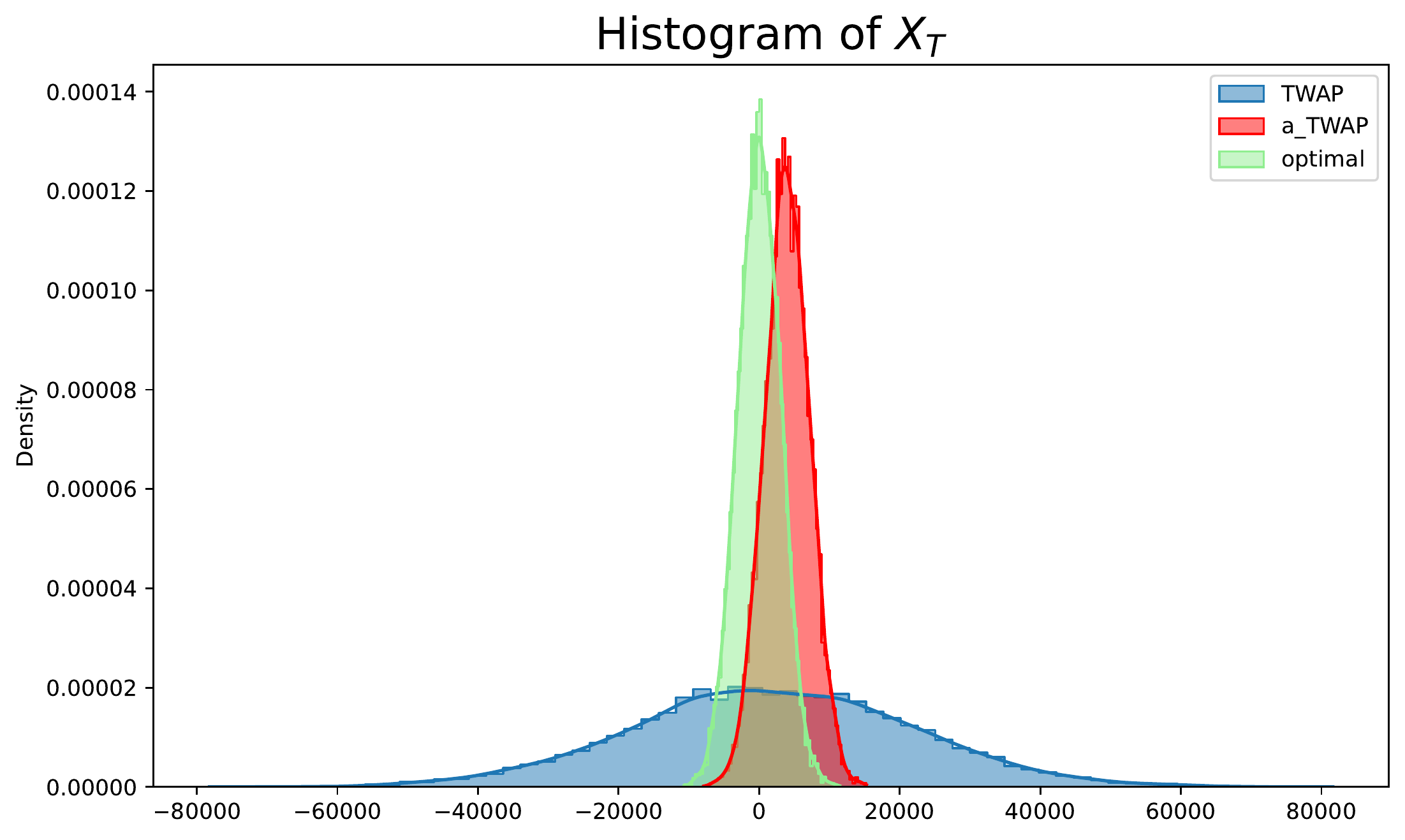}
\end{figure}

For a risk neutral trader, i.e., $\lambda=0$, Figures \ref{fig:case2} through \ref{fig:xT_hist_case2} exhibit respectively the expected remaining positions $X(\cdot)$ during liquidation, the histograms of objective functionals, as well as the histograms of the terminal shares $X(\cdot)$ prior to a final block trade. Solid lines indicate kernel density estimate for the histograms. The adapted TWAP in this setting reads
\begin{equation*}
v(t) = \frac{X(t)}{1 - t - 0.010005}.
\end{equation*}
In this case, we observe that, since the expected optimal trading trajectory is pretty close to TWAP strategy, the histograms from applying adapted TWAP and the optimal strategies are almost identical. Note that Figure \ref{fig:value_hist_case2} shows that the histogram associated with TWAP generates a distribution of performances that is severely left skewed, showing that the major loss w.r.t. the optimal strategy is driven by higher moment risk.
Likewise, in Figure \ref{fig:xT_hist_case2} we report the histograms of terminal position $X(T)$ prior to final block trade. Those for adapted TWAP and the optimal strategy are more concentrated than TWAP, again indicating that adapted strategies substantially reduce dispersion. Note also that the optimal strategy achieves a lower mean size of the final block liquidation size with respect to the adapted TWAP, thus proving that the optimal strategy produces a systematic reduction of the average final block trade, that is determined by properly taking into account the transient price impact determined by the market maker inventories' management.

\begin{figure}[b!]
\caption{Expected trading trajectory during the course of execution for risk averse trader. Optimal in red, TWAP in blue, and Almgren-Chriss in orange.\\} \label{fig:case1}
\includegraphics[height=6cm,width=9cm]{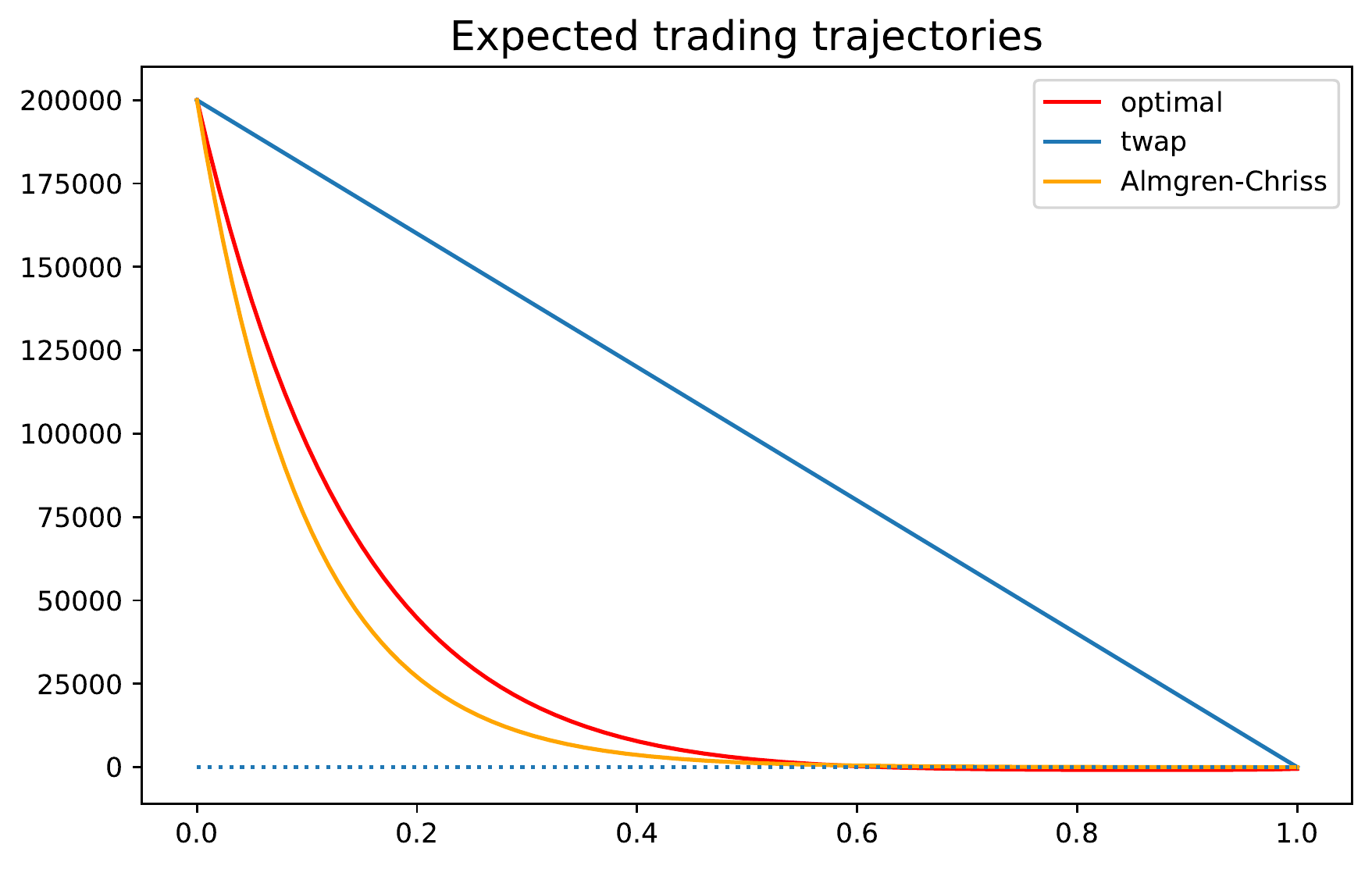}
\end{figure}

\begin{figure}[h!]
\caption{Histogram with kernel density estimate of objective functional for risk averse trader. Optimal in green, TWAP in blue, adapted TWAP in red, and Almgren-Chriss in orange.\\} \label{fig:value_hist}
\includegraphics[height=6cm,width=9cm]{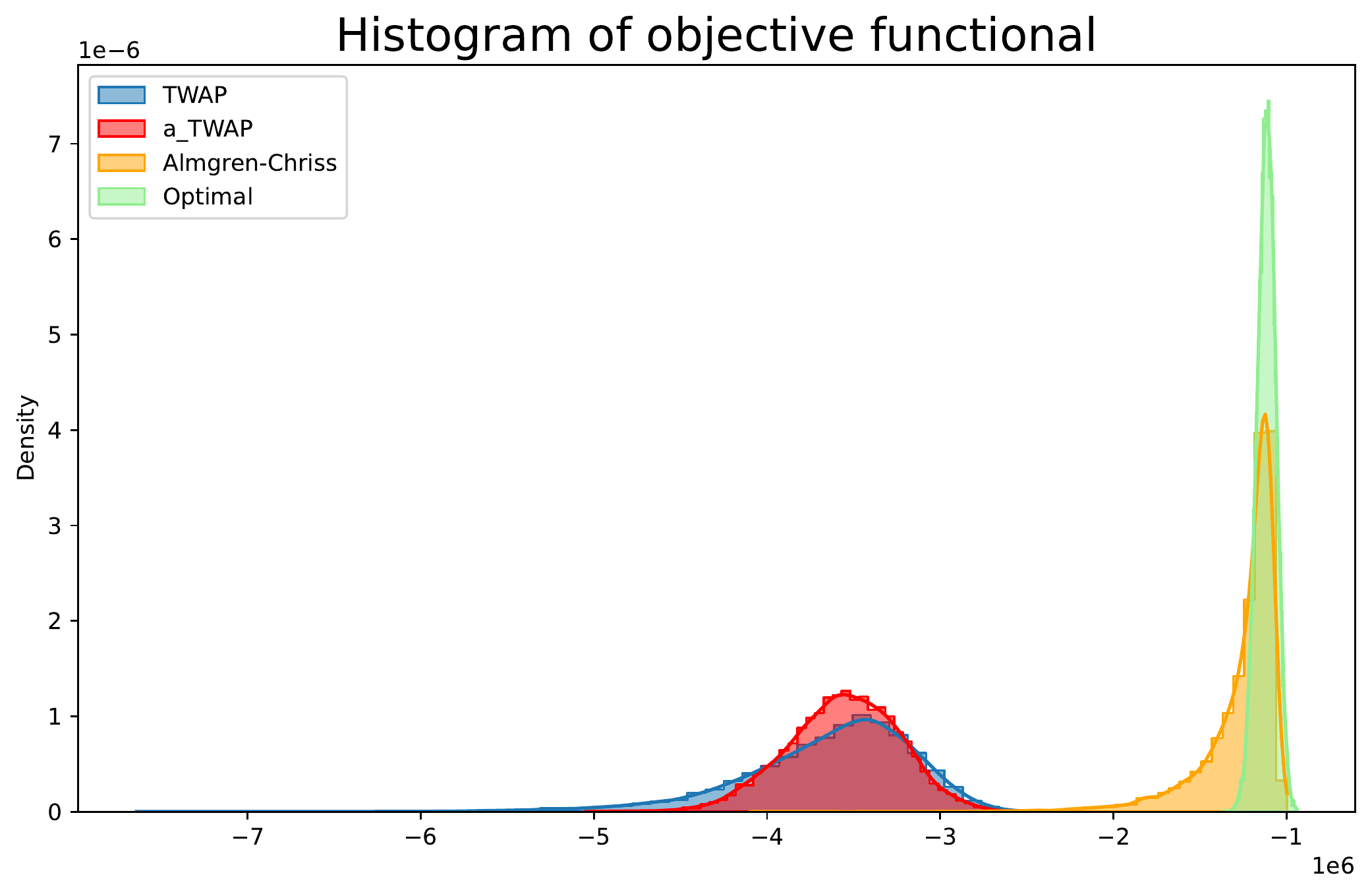}
\end{figure}

\begin{figure}[h!]
\caption{Histogram with kernel density estimate for terminal position $X_T$ to be liquidated by a block trade for risk averse trader. Optimal in green, TWAP in blue, adapted TWAP in red, and Almgren-Chriss in orange.\\} \label{fig:xT_hist}
\includegraphics[height=6cm,width=9cm]{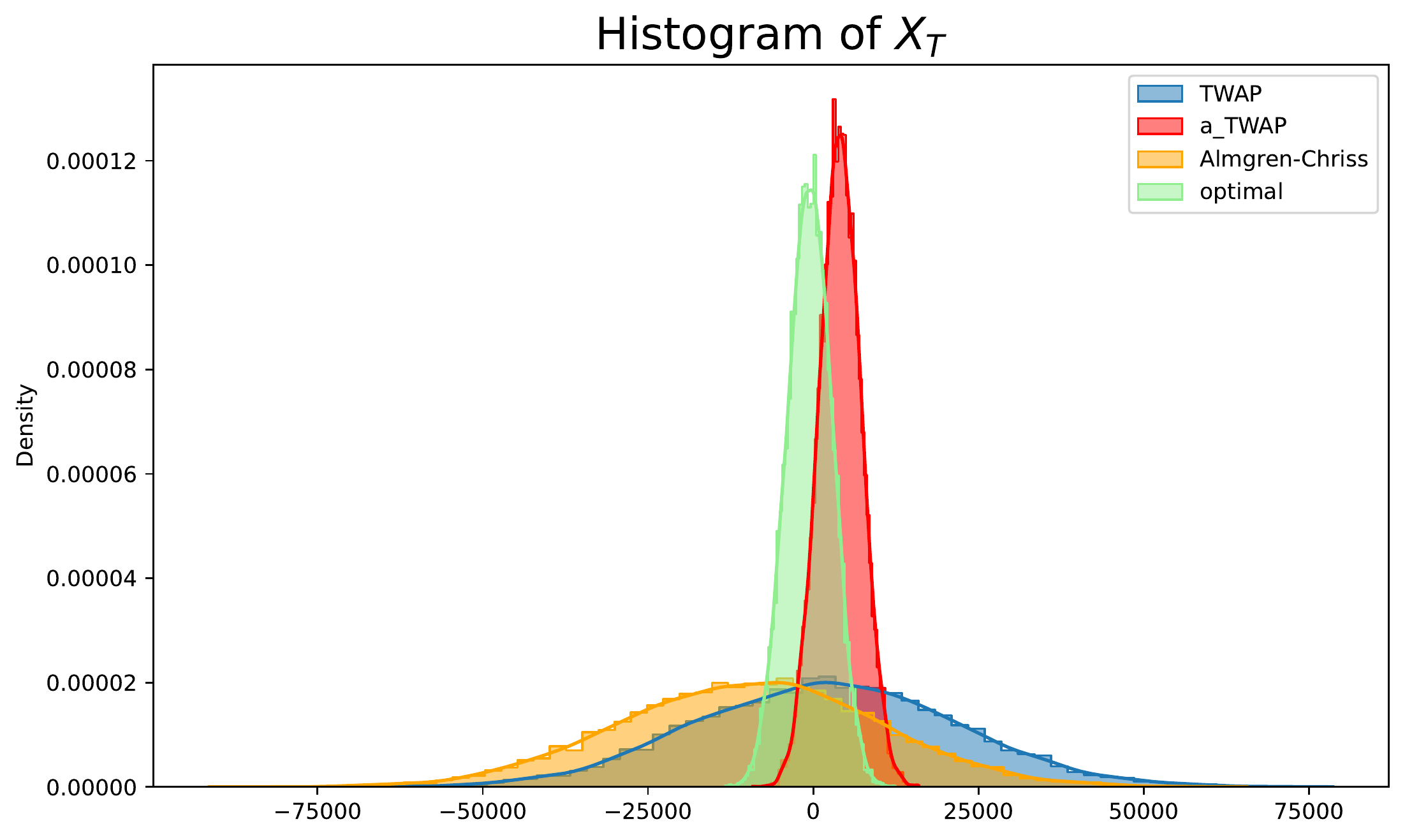}
\end{figure}

Then we set $\lambda = 0.001$ and consider a risk averse trader. Figures \ref{fig:case1} through \ref{fig:xT_hist} exhibit respectively the expected remaining positions $X(\cdot)$ during liquidation, the histograms of objective functionals, as well as the histograms of the terminal shares $X(\cdot)$ prior to a final block trade. Solid lines indicate kernel density estimates for the histograms. For reader's convenience, we recall that the Almgren-Chriss strategy (consisting of the number of units to be sold) is given by
\begin{equation*}
X(t) = x_{0} \frac{\sinh(\kappa(T - t))}{\sinh(\kappa T)},
\end{equation*}
where $\kappa := \sqrt{\frac{\lambda\sigma_S^{2}}\eta}$. In our context, $T = 1$ and $\kappa = 10$. The expected optimal trading strategies in Figures \ref{fig:case1} show that the optimal one is substantially deviating from TWAP type strategies and is closer to the Almgren and Chriss one. The histogram of the distribution of the achieved objective functional levels for different strategies in Figure \ref{fig:value_hist} shows the dramatic decline of TWAP-like strategies' performance and the high level of left skewness and higher moment risk generated also by the Almgren-Chriss one. Finally, looking at Figure \ref{fig:xT_hist} it is also evident that, despite the marginal change in the strategy,  the histograms of terminal position $X(T)$ prior to final block trade for adapted TWAP and the optimal strategies are more concentrated than TWAP and the Almgren-Chriss strategies, indicating that adapted strategies are able to hedge optimally risk arising from uncertainty of order fills.


\section{Conclusions} \label{sec:conclusion}

We introduced in this article a price impact model that takes into account a contribution representing the price pressure driven by market makers' inventories' risk. The resulting model is expected to be flexible enough to capture some well known stylized features of the empirically documented price response behavior during order execution and simple enough to be analytically tractable. The numerical illustration shows that the resulting optimal liquidation policy provides substantial performance improvements in relation to higher moment risk that emerges by analyzing the performance statistics.

Clearly, an effective characterization of the optimal policy based on public information about the market structure is still out of reach, in fact under a normal condition market maker inventories cannot be
observed. However it is worth observing that this one interesting direction of development of the current framework will consist in integrating this optimization approach within a partial information scheme. Then, in light of
the linear-quadratic nature of the optimal solution, it is easy to envisage the possibility to extend the model including also a procedure that filters inventories' size from publicly available information to achieve the goal of reproducing a realistic price impact function as an outcome of optimal behavior of all the market participants.


\section*{Acknowledgement}
We wish to thank Peter Carr for his valuable comments. We are also grateful to all participants of the following seminars and conferences: Brooklyn Quant Experience Lecture Series (February 27th, 2020: Tandon School of Engineering, New York University, USA), Mathematical Finance Seminar (April 15, 2021: Ritsumeikan University, Japan), Financial Engineering Seminar Series (March 25, 2021: Stevens Institute of Technology, USA), XXI Workshop of Quantitative Finance (January 27-29, 2020: Università degli studi di Napoli Parthenope, Italy), 10th General AMaMeF Conference (June 22-25, 2021: Università degli studi di Padova, Italy). The usual disclaimer applies.

Claudio Tebaldi acknowledges financial support by grant MIUR - PRIN Bando 2017 - prot. 2017TA7TYC.

Tai-Ho Wang is partially supported by the PSC-CUNY52 Award and the National Science Foundation of China grant 11971040.


\printbibliography


\appendix


\section{Convergence to Ornstein-Uhlenbeck process} \label{app:conv-ou}
We show in this appendix that the dynamics of market maker's inventory when following the approximately optimal quoting rule given in   eq.s \eqref{eq:delta} converges to an Ornstein-Uhlenbeck process in the limit as $h$ approaches zero.

Recall that $S^{b}$ and $S^{a}$ denote the bid and ask prices, respectively, and $\delta^{b}$ and $\delta^{a}$ the difference between the quotes and the reference price $S$, i.e., $\delta^{b} =:S-S^{b}$ and $\delta^{a}=:S^{a}-S$. The reference price $S$ is assumed following an arithmetic Brownian motion with drift $\mu \geq 0$ and volatility $\sigma > 0$. Let $N_t^b$ and $N_t^a$ be the Poisson processes that represent respectively the cumulative market sell and market buy orders up to time $t$. Thus, market maker's inventory $q_t$ at time $t$ is given by $q_t = N_t^b - N_t^a$. The arrival rates $\theta^a$ and $\theta^b$ for market orders are given in \eqref{eq:arrival-rates}.

Next, by scaling up the parameter $A$ (and abuse the use of notation) in \eqref{eq:arrival-rates} by
\begin{equation*}
A \longrightarrow \frac{A}{h^{2}}
\end{equation*}
for some fixed constant $A>0$ and scaling down the jump size of Poisson processes by $h$, i.e.,
\[
N_t^a \longrightarrow h N_t^a \quad \mbox{ and } \quad N_t^b \longrightarrow h N_t^b,
\]
the arrival rates becomes
\begin{equation*}
\theta^{b} (\delta^{b}) = \frac A{h^{2}} e^{-\kappa\delta^{a}}, \qquad \theta^{a}(\delta^{a}) = \frac A{h^{2}} e^{-\kappa \delta^{a}}
\end{equation*}
and market marker's inventory $q_t = h(N_t^b - N_t^a)$. We have that the approximately optimal quotes in \eqref{eq:delta} transform into
\begin{eqnarray*}
\delta^{b}(q) &=& \frac{1}{\nu} \ln\left( 1 + \frac{\nu}{\kappa} \right) + \left( q + \frac{1}{2} -\frac{\mu}{\nu \sigma^{2}} \right) \sqrt{\frac{\sigma^{2}\nu h^{2}}{2\kappa A} \left( 1 + \frac{\nu}{\kappa} \right)^{1+\frac{\kappa}{\nu}}}, \\
\delta^{a}(q) &=& \frac{1}{\nu} \ln\left( 1 + \frac{\nu}{\kappa} \right) +\left( -q + \frac{1}{2} +\frac{\mu}{\nu \sigma^{2}} \right) \sqrt{\frac{\sigma^{2}\nu h^{2}}{2\kappa A} \left( 1 + \frac{\nu}{\kappa} \right)^{1+\frac{\kappa}{\nu}}}.
\end{eqnarray*}
It follows that the infinitesimal generator $\mathcal A$ for market maker's inventory $q_t$ under the rule given above takes the form, for any given properly defined function $u$,
\begin{align*}
\mathcal A u(q) &= \theta^{b}(\delta^{b}(q))\left[u(q+h)-u(q)\right] + \theta^{a}(\delta^{a}(q))\left[u(q-h)-u(q)\right] \\
&\begin{aligned}
=& \frac{\theta^{b}(\delta^{b}(q)) + \theta^{a}(\delta^{a}(q))}{2} \left[u(q+h)-2u(q)+u(q-h) \right] \\[3pt]
& + \frac{\theta^{b}(\delta^{b}(q)) - \theta^{a}(\delta^{a}(q))}{2} \left[u(q+h)-u(q-h) \right] .
\end{aligned}
\end{align*}

We claim that the infinitesimal generator $\mathcal A$ converges to that of an Ornstein-Uhlenbeck process as $h \to 0^{+}$.
Indeed, setting
\begin{eqnarray*}
c_{1} &=& \frac A2 \left(1 + \frac{\nu}{\kappa} \right)^{-\frac{\kappa}{\nu}}, \\[6pt]
c_{2} &=& \sqrt{\frac{\sigma^{2} \nu}{2\kappa A} \left(1 + \frac{\nu}{\kappa} \right)^{1+\frac{\kappa}{\nu}}},
\end{eqnarray*}
we rewrite $\mathcal A$ as
\small{
\begin{multline*}
\mathcal A u(q) \\
= c_{1} \exp\left\{-\frac{khc_{2}}{2}\right\} \left[\exp\left\{-khc_{2} \left(q-\frac{\mu}{\nu \sigma_{S}^{2}}\right)\right\} + \exp\left\{khc_{2}\left(q-\frac{\mu}{\nu \sigma_{S}^{2}}\right)\right\}\right]
 \frac{u(q+h)-2u(q)+u(q-h)}{h^{2}} \\
+ 2c_{1} \exp\left\{-\frac{khc_{2}}{2}\right\} \left[\frac{\exp\left\{-khc_{2} \left(q-\frac{\mu}{\nu \sigma_{S}^{2}}\right)\right\} -\exp\left\{khc_{2}\left(q-\frac{\mu}{\nu \sigma_{S}^{2}}\right)\right\}}{h}\right] \left(\frac{u(q+h)-u(q-h) }{2h} \right).
\end{multline*}
}
\normalsize
It follows by taking the limit $h \to 0^+$ on both side of the last expression that
\begin{equation*}
\lim_{h \to 0^+}\mathcal A u(q) =
2c_{1} u^{\prime\prime}(q) + 4c_{1}c_{2}\kappa\left( \frac{\mu}{\nu \sigma^{2}} - q \right)u^{\prime}(q)
\end{equation*}
if $u$ is twice differentiable. We conclude that the limiting infinitesimal generator on the right hand side of the previous expression is of an Ornstein-Uhlenbeck process with mean reversion rate $2c_{1}c_{2}\kappa$, long term mean $\frac{\mu}{\nu \sigma_{S}^{2}}$ and volatility $2\sqrt{c_{1}}$.


\section{Proof of Corollary \ref{cor:cor1}} \label{app:proof-cor1}

Note that in this case the system of ODEs reduce respectively to
\begin{multline} \label{eq:M-odes-reduced}
\frac{d}{dt} \left[\begin{array}{cc}
M_{11} & M_{12} \\
M_{21} & M_{22}
\end{array}\right] \\
= \left[\begin{array}{cc}
\bs\Theta M_{11} - \frac\phi{2\teta} \bs\nu M_{21} - \frac{\phi^2}{4\teta} \bs\nu\bs\nu' N_{11} - \frac{\phi\widetilde{\xi}}{4\teta} \bs\nu N_{21} & \bs\Theta M_{12} - \frac\phi{2\teta} \bs\nu M_{22} - \frac{\phi^2}{4\teta} \bs\nu\bs\nu' N_{12} - \frac{\phi\widetilde{\xi}}{4\teta} \bs\nu N_{22} \\
-\frac{\widetilde{\xi}}{2\teta} M_{21} - \frac{\phi\widetilde{\xi}}{4\teta}\bs\nu' N_{11} + \left(-\frac{\widetilde{\xi}^2}{4\teta} + \psi\right)N_{21} & -\frac{\widetilde{\xi}}{2\teta} M_{22} - \frac{\phi\widetilde{\xi}}{4\teta}\bs\nu' N_{12} + \left(-\frac{\widetilde{\xi}^2}{4\teta} + \psi\right)N_{22}
\end{array}\right]
\end{multline}
and
\begin{equation}\label{eq:N-odes-reduced}
\frac{d}{dt} \left[\begin{array}{cc}
N_{11} & N_{12} \\
N_{21} & N_{22}
\end{array}\right]
= \left[\begin{array}{cc}
-\bs\Theta N_{11} & -\bs\Theta N_{12} \\
\frac1\teta M_{21} + \frac\phi{2\teta} \bs\nu' N_{11} + \frac{\widetilde{\xi}}{2\teta}N_{21} & \frac1\teta M_{22} + \frac\phi{2\teta} \bs\nu' N_{12} + \frac{\widetilde{\xi}}{2\teta}N_{22}
\end{array}\right].
\end{equation}
We have, from \eqref{eq:N-odes-reduced} and taking into account the terminal conditions $N_{11}(T) = \bs I$ and $N_{12}(T) = 0$ , that
\begin{equation*}
N_{11} = e^{(T - u)\Theta}, \qquad N_{12} = \bs 0.
\end{equation*}
Next notice that $M_{22}$ and $N_{22}$ satisfy the following coupled ODEs
\begin{align*}
\dot M_{22} &= -\frac{\widetilde{\xi}}{2\teta} M_{22} + \left(-\frac{\widetilde{\xi}^2}{4\teta} + \psi\right)N_{22}, \\
\dot N_{22} &= \frac1\teta M_{22} + \frac{\widetilde{\xi}}{2\teta}N_{22}.
\end{align*}

Define $\tM_{22} = M_{22} + \frac{\widetilde{\xi}}2 N_{22}$.
Since
\begin{equation*}
\dot M_{22} = \dot\tM_{22} - \frac{\widetilde{\xi}}2 \dot N_{22},
\end{equation*}
We have
\begin{align*}
\dot M_{22} &= -\frac{\widetilde{\xi}}{2\teta} \tM_{22} + \psi N_{22}, \\
\dot N_{22} &= \frac1\teta \tM_{22}
\end{align*}
and
\begin{align*}
\dot\tM_{22} &= \dot M_{22} + \frac{\widetilde{\xi}}2 \dot N_{22} \\
&= -\frac{\widetilde{\xi}}{2\teta} \tM_{22} + \psi N_{22} + \frac{\widetilde{\xi}}{2\teta} \tM_{22} \\
&= \psi N_{22}
\end{align*}
Hence, $\tM_{22}$ and $N_{22}$ satisfy
\begin{align*}
\dot \tM_{22} &= \psi N_{22} \\
\dot N_{22} &= \frac1\teta \tM_{22} .
\end{align*}
Taking into the account the terminal conditions $\tM_{22}(T) = \frac{\gamma + \widetilde{\xi}}2 - \beta$ and $N_{22}(T) = 1$, we obtain the solutions for $\tM_{22}$ and $N_{22}$ as
\begin{align*}
\tM_{22} &= \left(\frac{\gamma+\widetilde{\xi}}2 - \beta\right) \cosh\left(\sqrt{\frac\psi\teta}\left\{T - u\right\}\right) - \sqrt{\psi\teta}\sinh\left( \sqrt{\frac\psi\teta}\left\{T - u\right\}\right), \\
N_{22} &= \cosh\left(\sqrt{\frac\psi\teta}\left\{T - u\right\}\right) - \frac1{\sqrt{\psi\teta}} \left(\frac{\gamma + \widetilde{\xi}}2 - \beta\right) \sinh\left( \sqrt{\frac\psi\teta}\left\{T - u\right\}\right).
\end{align*}
By using the notation $\talpha$ defined in \eqref{eq:tilde-alpha}, we further rewrite $\tM_{22}$ and $N_{22}$ as
\begin{eqnarray*}
&& \tM_{22}
= -\sqrt{\left(\frac{\gamma+\widetilde{\xi}}2 - \beta\right)^2 - \psi\teta} \; \cosh\left(\sqrt{\frac\psi\teta}\left\{T - u + \talpha\right\}\right), \\
&& N_{22} = \frac{\sinh\left(\sqrt{\frac\psi\teta}\left\{T - u + \talpha\right\}\right)}{\sinh\left(\sqrt{\frac\psi\teta}\talpha\right)}.
\end{eqnarray*}
It follows that
\begin{equation*}
\frac{\tM_{22}}{N_{22}} = -\sqrt{\psi\teta} \coth\left(\sqrt{\frac\psi\teta}\left\{T - u + \talpha\right\}\right).
\end{equation*}
We solve $M_{12}$ as follows. Note that, from \eqref{eq:M-odes-reduced} and since $N_{12} = \bs 0$, $M_{12}$ satisfies
\begin{align*}
&\dot M_{12} = \bs\Theta M_{12} - \frac\phi{2\teta} \bs\nu M_{22} - \frac{\phi\widetilde{\xi}}{4\teta} \bs\nu N_{22} = \bs\Theta M_{12} - \frac\phi{2\teta} \bs\nu \tM_{22} \\
&\Longrightarrow \quad \frac{d}{dt}\left\{e^{-u\bs\Theta} M_{12}\right\} = -\frac{\phi}{2\teta} e^{-u\bs\Theta} \tM_{22} \bs\nu  \\
&\Longrightarrow \quad e^{-T\bs\Theta}M_{12}(T) - e^{-u\bs\Theta} M_{12} = -\frac{\phi}{2\teta} \int_u^T e^{-s\bs\Theta}\tM_{22}(s) ds \bs\nu \\
&\Longrightarrow\quad M_{12} = -\frac\phi2 e^{(u-T)\bs\Theta}\bs\nu + \frac{\phi}{2\teta} \int_u^T e^{(u-s)\bs\Theta}\tM_{22}(s) ds \bs\nu ,
\end{align*}
since $M_{12}(T) = -\frac\phi2\bs\nu$. Therefore,
\begin{equation*}
M_{12} = -\frac\phi2 \left\{e^{(u-T)\bs\Theta}\bs I + \frac{\zeta}{\sinh(\zeta\talpha)} \int_u^T e^{(u-s)\bs\Theta} \cosh\left(\zeta\left\{T - s + \talpha\right\}\right) ds \right\} \bs\nu,
\end{equation*}
where $\zeta = \sqrt{\frac\psi\teta}$ for notational simplicity.
We evaluate the integral on the right hand side as follows. Since
\begin{align*}
&\int_u^T e^{(u-s)\bs\Theta} \cosh\left\{\zeta(T - s + \talpha)\right\} ds \\
&= -\frac1\zeta \left(\bs I - \frac1{\zeta^2}\bs\Theta^2\right)^{-1} e^{(u-T)\bs\Theta}\sinh\left\{\zeta \talpha \right\} + \frac1{\zeta^2} \left(\bs I - \frac1{\zeta^2}\bs\Theta^2\right)^{-1} \bs\Theta e^{(u-T)\bs\Theta} \cosh(\zeta\talpha) \\
&\quad +\frac{1}\zeta \left(\bs I - \frac{1}{\zeta^2}\bs\Theta^2\right)^{-1} \sinh\left\{\zeta(T - u + \talpha)\right\} - \frac1{\zeta^2} \left(\bs I - \frac{1}{\zeta^2}\bs\Theta^2\right)^{-1} \bs\Theta \cosh(\zeta(T - u + \talpha)),
\end{align*}
we obtain that
\begin{eqnarray*}
M_{12} &=& -\frac\phi2 \left\{e^{(u-T)\bs\Theta} \bs I + \frac{\zeta}{\sinh(\zeta\talpha)} \int_u^T e^{(u-s)\bs\Theta} \cosh\left(\zeta\left\{T - s + \talpha\right\}\right) ds \right\} \bs\nu \\
&=& -\frac\phi2 \left\{ e^{(u-T)\bs\Theta}\bs I \right. \\
&& - \left(\bs I - \frac1{\zeta^2}\bs\Theta^2\right)^{-1} e^{(u-T)\bs\Theta} + \frac1{\zeta} \left(\bs I - \frac1{\zeta^2}\bs\Theta^2\right)^{-1} \bs\Theta e^{(u-T)\bs\Theta} \coth(\zeta\talpha) \\
&& \left. + \left(\bs I - \frac1{\zeta^2}\bs\Theta^2\right)^{-1} \frac{\sinh\left\{\zeta(T - u + \talpha)\right\}}{\sinh(\zeta\talpha)} - \frac1{\zeta} \left(\bs I - \frac1{\zeta^2}\bs\Theta^2\right)^{-1} \bs\Theta \frac{\cosh(\zeta(T - u + \talpha))}{\sinh(\zeta\talpha)} \right\} \bs\nu \\
&=& -\frac\phi2 \left\{e^{(u-T)\bs\Theta} \bs I + \frac{\zeta}{\sinh(\zeta\talpha)} \int_u^T e^{(u-s)\bs\Theta} \cosh\left(\zeta\left\{T - s + \talpha\right\}\right) ds \right\} \bs\nu \\
&=& -\frac\phi2 \left\{ e^{(u-T)\bs\Theta}\left[ \bs I - \left(\bs I - \frac1{\zeta^2}\bs\Theta^2\right)^{-1} + \frac1{\zeta} \left(\bs I - \frac1{\zeta^2}\bs\Theta^2\right)^{-1} \bs\Theta \coth(\zeta\talpha) \right] \right. \\
&& \left. + \left(\bs I - \frac1{\zeta^2}\bs\Theta^2\right)^{-1} \frac{\sinh\left\{\zeta(T - u + \talpha)\right\}}{\sinh(\zeta\talpha)} - \frac1{\zeta} \left(\bs I - \frac1{\zeta^2}\bs\Theta^2\right)^{-1} \bs\Theta \frac{\cosh(\zeta(T - u + \talpha))}{\sinh(\zeta\talpha)} \right\} \bs\nu
\end{eqnarray*}
and
\begin{align*}
&\int_{u}^{T} M_{12}(s) ds \\
&= -\frac\phi2 \left\{ \bs\Theta^{-1} \left[ \bs I - e^{(u-T)\bs\Theta} \right] \left[ \bs I - \left(\bs I - \frac1{\zeta^2}\bs\Theta^2\right)^{-1} + \frac1{\zeta} \left(\bs I - \frac1{\zeta^2}\bs\Theta^2\right)^{-1} \bs\Theta \coth(\zeta\talpha) \right] \right. \\
&\quad \left. +\frac1\zeta \left(\bs I - \frac{1}{\zeta^2}\bs\Theta^2\right)^{-1} \left[-\coth(\zeta\talpha) + \frac{\cosh\left\{\zeta(T - u + \talpha)\right\}}{\sinh(\zeta\talpha)} \right] \right. \\
&\quad \left. -\frac1{\zeta^2} \left(\bs I -\frac{1}{\zeta^2}\bs\Theta^2\right)^{-1} \bs\Theta \left[ -1 + \frac{\sinh(\zeta(T - u + \talpha))}{\sinh(\zeta\talpha)} \right] \right\} \bs\nu
\end{align*}
Thus,
\begin{align*}
&\frac{2}{N_{22}} \int_u^T M_{12}(s) ds \\
&= -\phi \left\{ \frac{\sinh(\zeta\talpha)}{\sinh(\zeta(T - u + \talpha))} \bs\Theta^{-1} \left[ \bs I - e^{(u-T)\bs\Theta} \right] \left[ \bs I - \left(\bs I - \frac1{\zeta^2}\bs\Theta^2\right)^{-1} \right.\right. \\
&\quad \left. + \frac{1}{\zeta} \left(\bs I - \frac1{\zeta^2}\bs\Theta^2\right)^{-1} \bs\Theta \coth(\zeta\talpha) \right] \\
&\quad \left.+\frac{1}\zeta \left(\bs I - \frac1{\zeta^2}\bs\Theta^2\right)^{-1} \left[-\frac{\cosh(\zeta\talpha)}{\sinh(\zeta(T - u + \talpha))} + \coth\left\{\zeta(T - u + \talpha)\right\} \right] \right. \\
&\quad \left. - \frac1{\zeta^2} \left(\bs I - \frac1{\zeta^2}\bs\Theta^2\right)^{-1} \bs\Theta \left[ 1 - \frac{\sinh(\zeta\talpha)}{\sinh(\zeta(T - u + \talpha))} \right] \right\} \bs\nu
\end{align*}
Hence,
\begin{align*}
&\frac2{N_{22}} \int_u^T M_{12}'(s) ds \bs\Theta\bs\bq_0 \\
&= -\phi \bs\nu' \left\{ \frac{\sinh(\zeta\talpha)}{\sinh(\zeta(T - u + \talpha))} \left[ \bs I -e^{(u-T)\bs\Theta} \right] \left[ \bs I - \left(\bs I -\frac1{\zeta^2}\bs\Theta^2\right)^{-1} \right.\right. \\
&\quad +\left.\frac{1}{\zeta} \left(\bs I - \frac{1}{\zeta^2}\bs\Theta^2\right)^{-1} \bs\Theta \coth(\zeta\talpha) \right] \\
&\quad \left. + \frac1\zeta \left(\bs I - \frac1{\zeta^2}\bs\Theta^2\right)^{-1} \bs\Theta \left[-\frac{\cosh(\zeta\talpha)}{\sinh(\zeta(T - u + \talpha))} + \coth\left\{\zeta(T - u + \talpha)\right\} \right] \right. \\
&\quad \left. -\frac1{\zeta^2} \left(\bs I - \frac1{\zeta^2}\bs\Theta^2\right)^{-1} \bs\Theta^2 \left[ 1 - \frac{\sinh(\zeta\talpha)}{\sinh(\zeta(T - u + \talpha))} \right] \right\} \bs\bq_0
\end{align*}
Therefore,
\begin{align*}
\frac{M_{12}}{N_{22}} =& -\frac\phi2 \left\{ \frac{e^{(u-T)\bs\Theta}}{N_{22}} \left[ \bs I - \left(\bs I - \frac1{\zeta^2}\bs\Theta^2\right)^{-1} + \frac1{\zeta} \left(\bs I - \frac1{\zeta^2}\bs\Theta^2\right)^{-1} \bs\Theta \coth(\zeta\talpha) \right] \right. \\
&\quad \left. +\left(\bs I - \frac1{\zeta^2}\bs\Theta^2\right)^{-1} - \frac1{\zeta} \left(\bs I - \frac1{\zeta^2}\bs\Theta^2\right)^{-1} \bs\Theta \coth(\zeta(T - u + \talpha)) \right\} \bs\nu.
\end{align*}
Since
\[
N^{-1} = \left[\begin{array}{cc} N_{11}^{-1} & \bs 0 \\ -\frac1{N_{22}} N_{21} N_{11}^{-1} & \frac1{N_{22}} \end{array}\right]
\]
we have
\begin{equation*}
\begin{aligned}
\bR &= M N^{-1} \\
&= \left[\begin{array}{cc} M_{11} & M_{12} \\ M_{21} & M_{22} \end{array}\right] \,
\left[\begin{array}{cc} N_{11}^{-1} & \bs 0 \\ N_{21} & \dfrac{1}{N_{22}} \end{array}\right] \\
&= \left[\begin{array}{cc} M_{11} N_{11}^{-1} + M_{12} N_{21} & \dfrac{M_{12}}{N_{22}} \\ \cdots & \dfrac{\tM_{22}}{N_{22}} - \dfrac{\widetilde{\xi}}2 \end{array}\right].
\end{aligned}
\end{equation*}
Thus,
\begin{equation*}
\begin{aligned}
R_{12} &= \dfrac{M_{12}}{N_{22}} \\
&= -\frac{\phi}{2} \left\{ \frac{e^{(u-T)\bs\Theta}}{N_{22}} \left[ \bs I - \left(\bs I - \frac1{\zeta^2}\bs\Theta^2\right)^{-1} + \frac1{\zeta} \left(\bs I - \frac1{\zeta^2}\bs\Theta^2\right)^{-1} \bs\Theta \coth(\zeta\talpha) \right] \right. \\
&\quad \left. + \left(\bs I - \frac{1}{\zeta^2}\bs\Theta^2\right)^{-1} - \frac1{\zeta} \left(\bs I - \frac1{\zeta^2}\bs\Theta^2\right)^{-1} \bs\Theta \coth(\zeta(T - u + \talpha)) \right\} \bs\nu.
\end{aligned}
\end{equation*}

The equation \eqref{eq:linear-term-n} for $\bs r$ becomes
\[
\dot \br + \bA'\br + 2 \bR \bb + \mu\be + \frac1{\widetilde\eta}\left(\bR \ba \ba' + \bk\ba'\right) \br = 0
\]
We shall focus on the term $r_{n+1}$ only since it is the term that is relevant to the determination of the optimal trading rate
\begin{align*}
&\dot r_{n+1} + 2 R_{21}\bs\Theta\bs\bq_0 + \mu + \frac1{\widetilde\eta} \left(R_{22} + \frac{\widetilde{\xi}}2\right) r_{n+1} = 0 \\
&\Longrightarrow \quad \dot r_{n+1} + 2 R_{21}\bs\Theta\bs\bq_0 + \mu + \frac1{\widetilde\eta} \frac{\tM_{22}}{N_{22}} r_{n+1} = 0 \\
&\Longrightarrow \quad \dot r_{n+1} + 2 R_{21}\bs\Theta\bs\bq_0 + \mu + \frac{\dot N_{22}}{N_{22}} r_{n+1} = 0 \\
&\Longrightarrow \quad N_{22} \dot r_{n+1} + \dot N_{22} r_{n+1} = -2 M_{12}'\bs\Theta\bs\bq_0 - \mu N_{22} \\
&\Longrightarrow \quad -N_{22} r_{n+1} = -2 \int_u^T M_{12}'(s) ds\bs\Theta\bs\bq_0 - \mu \int_u^T N_{22}(s) ds \\
&\Longrightarrow \quad r_{n+1} = \frac2{N_{22}} \int_u^T M_{12}'(s) ds\bs\Theta\bs\bq_0 + \frac\mu{N_{22}} \int_u^T N_{22}(s) ds.
\end{align*}

Note that since $N_{22} = \dfrac{1}{\psi} \dot\tM_{22}$, we have
\begin{align*}
&-\frac1{2\teta}\frac\mu{N_{22}}\int_u^T N_{22}(s) ds = -\frac\mu{2\psi\teta} \left(\frac{\frac{\gamma + \widetilde{\xi}}2- \beta}{N_{22}} - \frac{\tM_{22}}{N_{22}}\right) \\
&= -\frac\mu{2\psi\teta} \left[ \frac{\frac{\gamma + \widetilde{\xi}}2- \beta}{N_{22}} + \sqrt{\psi\teta} \coth(\zeta(T - u + \talpha)) \right] \\
&= -\frac\mu{2\sqrt{\psi\teta}} \left[ \frac{\frac{\gamma + \widetilde{\xi}}2- \beta}{\sqrt{\psi\teta}} \frac{\sinh(\zeta\talpha)}{\sinh(\zeta(T - u + \talpha))} + \coth(\zeta(T - u + \talpha)) \right] \\
&= \frac\mu{2\sqrt{\psi\teta}} \left[ \frac{\cosh(\zeta\talpha)}{\sinh(\zeta(T - u + \talpha))} - \coth(\zeta(T - u + \talpha)) \right].
\end{align*}

Finally, the optimal trading rate $v^\star$ in this case reads
\begin{align*}
v^{\star}(u) &= \frac1{2\widetilde{\eta}} \left\{ 2 (\bk + \bR(u) \ba)'\bs{X}_{G}(u) + \bs{a}' \br(u) \right\} \\
&= -\frac\phi{2\teta} \bs\nu'\bs{Q}_{G}(u) -\frac{\widetilde{\xi}}{2\widetilde\eta} X_{G}(u) - \frac1\teta (R_{12}(u)'\bs{Q}_{G}(u) + R_{22}(u) X_{G}(u)) - \frac1{2\teta}r_{n+1} \\
&= -\frac\phi{2\teta} \bs\nu'\bs{Q}_{G}(u) - \frac1\teta \frac{\tM_{22}}{N_{22}} X_{G}(u) - \frac1\teta \frac{M_{12}'}{N_{22}}\bs{Q}_{G}(u) - \frac1{2\teta}r_{n+1} \\
&= -\frac\phi{2\teta} \bs\nu'\bs{Q}_{G}(u) + \sqrt{\frac\psi\teta} \coth\left(\sqrt{\frac\psi\teta}\left\{T - u + \talpha\right\}\right) X_{G}(u) \\
&\quad +\frac\phi{2\teta} \bs\nu' \left\{ \frac{e^{(u-T)\bs\Theta}}{N_{22}} \left[ \bs I - \left(\bs I - \frac1{\zeta^2}\bs\Theta^2\right)^{-1} + \frac1{\zeta} \left(\bs I - \frac1{\zeta^2}\bs\Theta^2\right)^{-1} \bs\Theta \coth(\zeta\talpha) \right] \right. \\
&\quad \left. + \left(\bs I - \frac{1}{\zeta^2}\bs\Theta^2\right)^{-1} - \frac1{\zeta} \left(\bs I - \frac1{\zeta^2}\bs\Theta^2\right)^{-1} \bs\Theta \coth(\zeta(T - u + \talpha)) \right\} \bs{Q}_{G}(u) \\
&\quad +\frac{\phi}{2\teta} \bs\nu' \left\{ \frac{\sinh(\zeta\talpha)}{\sinh(\zeta(T - u + \talpha))} \left[ \bs I -e^{(u-T)\bs\Theta} \right] \left[ \bs I - \left(\bs I -\frac{1}{\zeta^2}\bs\Theta^2\right)^{-1} \right.\right.\\
&\quad \left.+ \frac{1}{\zeta} \left(\bs I - \frac{1}{\zeta^2}\bs\Theta^2\right)^{-1} \bs\Theta \coth(\zeta\talpha) \right] \\
&\quad \left. + \frac1\zeta \left(\bs I - \frac{1}{\zeta^2}\bs\Theta^2\right)^{-1} \bs\Theta \left[-\frac{\cosh(\zeta\talpha)}{\sinh(\zeta(T - u + \talpha))} + \coth\left\{\zeta(T - u + \talpha)\right\} \right] \right. \\
&\quad \left. -\frac{1}{\zeta^2} \left(\bs I - \frac{1}{\zeta^2}\bs\Theta^2\right)^{-1} \bs\Theta^2 \left[ 1 -\frac{\sinh(\zeta\talpha)}{\sinh(\zeta(T - u + \talpha))} \right] \right\} \bs\bq_0 \\
&\quad + \frac\mu{2\sqrt{\psi\teta}} \left\{ \frac{\cosh(\zeta\talpha)}{\sinh(\zeta(T - u + \talpha))} - \coth(\zeta(T - u + \talpha)) \right\} \\
&= \sqrt{\frac\psi\teta} \coth\left(\sqrt{\frac\psi\teta}\left\{T - u + \talpha\right\}\right) X_{G}(u)\\
&\quad +\frac\mu{2\sqrt{\psi\teta}} \left[ \frac{\cosh(\zeta\talpha)}{\sinh(\zeta(T - u + \talpha))} - \coth(\zeta(T - u + \talpha)) \right] \\
&\quad -\frac\phi{2\teta} \bs\nu'\bs{Q}_{G}(u) + \frac\phi{2\teta} \bs\nu' \left[ \bs\Lambda^{u}(T) \bs{Q}_{G}(u) + \bs\Lambda^{u}_{0}(T) \bs \bq_0\right],
\end{align*}
where $\bs\Lambda^{u}(T)$ and $\bs\Lambda^{u}_{0}(T)$ are given in \eqref{eq:Pi-t} and \eqref{eq:Pi0-t}, respectively.
\end{document}